\documentclass[prl,reprint,twocolumn,showpacs,superscriptaddress]{revtex4-2}
\usepackage{graphicx}
\usepackage{dcolumn}
\usepackage{bm}
\usepackage{graphicx}
\usepackage{epsfig}
\usepackage{epsf}
\usepackage{amssymb}
\usepackage{amsmath,empheq,cases}
\usepackage{amsthm}
\usepackage{xcolor}
\usepackage{mathtools}
\usepackage{multirow}
\usepackage[colorlinks=true,linkcolor=blue,citecolor=blue,urlcolor=blue,pdfauthor={ },pdftitle={ },pdfsubject={ },pdfkeywords={ }]{hyperref}
\usepackage{pgfplots}
\usepackage{lipsum}

\usepackage{times}

\usepackage{color}
\definecolor{dred}{rgb}{.8,0.2,.2}
\definecolor{ddred}{rgb}{.8,0.5,.5}
\definecolor{dblue}{rgb}{.2,0.2,.8}
\definecolor{dgreen}{rgb}{.2,0.5,.2}

\usepackage{tikz}
\definecolor{c2}{RGB}{153,51,51}

\usepackage[colorinlistoftodos,bordercolor=red!50,backgroundcolor=red!10,linecolor=red!50,textsize=tiny]{todonotes}
\usepackage{changes}

\newcommand\redsout{\bgroup\markoverwith{\textcolor{red}{\rule[0.5ex]{2pt}{1pt}}}\ULon}

\begin{document}
\title{Experimental Realization of Self-Contained Quantum Refrigeration}
\author{Keyi Huang}

\affiliation{Shenzhen Institute for Quantum Science and Engineering and Department of Physics, Southern
University of Science and Technology, Shenzhen 518055, China}	

\author{Cheng Xi}
\affiliation{Shenzhen Institute for Quantum Science and Engineering and Department of Physics, Southern University of Science and Technology, Shenzhen 518055, China}
\affiliation{Department of Physics, City University of Hong Kong, Tat Chee Avenue, Kowloon, Hong Kong SAR, China}

\author{Xinyue Long}
\affiliation{Shenzhen Institute for Quantum Science and Engineering and Department of Physics, Southern University of Science and Technology, Shenzhen 518055, China}

\author{Hongfeng Liu}
\affiliation{Shenzhen Institute for Quantum Science and Engineering and Department of Physics, Southern University of Science and Technology, Shenzhen 518055, China}

\author{Yu-ang Fan}
\affiliation{Shenzhen Institute for Quantum Science and Engineering and Department of Physics, Southern University of Science and Technology, Shenzhen 518055, China}

\author{Xiangyu Wang}
\affiliation{Shenzhen Institute for Quantum Science and Engineering and Department of Physics, Southern University of Science and Technology, Shenzhen 518055, China}

\author{Yuxuan Zheng}
\affiliation{Shenzhen Institute for Quantum Science and Engineering and Department of Physics, Southern University of Science and Technology, Shenzhen 518055, China}

\author{Yufang Feng}
\affiliation{Shenzhen Institute for Quantum Science and Engineering and Department of Physics, Southern University of Science and Technology, Shenzhen 518055, China}

\author{Xinfang Nie}
\email{niexf@sustech.edu.cn}
\affiliation{Shenzhen Institute for Quantum Science and Engineering and Department of Physics, Southern University of Science and Technology, Shenzhen 518055, China}
\affiliation{Quantum Science Center of Guangdong-Hong Kong-Macao Greater Bay Area, Shenzhen 518045, China}

\author{Dawei Lu}
\email{ludw@sustech.edu.cn}
\affiliation{Shenzhen Institute for Quantum Science and Engineering and Department of Physics, Southern University of Science and Technology, Shenzhen 518055, China}
\affiliation{Quantum Science Center of Guangdong-Hong Kong-Macao Greater Bay Area, Shenzhen 518045, China}
\affiliation{International Quantum Academy, Shenzhen 518055, China}

\date{\today}

\begin{abstract}
A fundamental challenge in quantum thermodynamics is the exploration of inherent dimensional constraints in thermodynamic machines. In the context of two-level systems, the most compact refrigerator necessitates the involvement of three entities, operating under self-contained conditions that preclude the use of external work sources. Here, we build such a smallest refrigerator using a nuclear spin system, where three distinct two-level carbon-13 nuclei in the same molecule are involved to facilitate the refrigeration process. The self-contained feature enables it to operate without relying on net external work, and the unique mechanism sets this refrigerator apart from its classical counterparts. We evaluate its performance under varying conditions and systematically scrutinize the cooling constraints across a spectrum of scenarios, which sheds light on the interplay between quantum information and thermodynamics.
\end{abstract}

\maketitle

\emph{Introduction.}---
Ever since Carnot's pioneering exploration of steam engines, which laid the foundation for thermodynamics, how to efficiently extract additional heat from an already cooled system has remained a central concern.  The progress in cooling techniques  has unlocked doors to extraordinary phenomena that emerge at very low temperatures~\cite{combescot2022superconductivity,PhysRevLett.75.3969,doi:10.1126/science.269.5221.198}, and has played a pivotal role in advancing the engineering of quantum systems~\cite{doi:10.1126/science.1134008,Tan2017,Nakamura1999,PhysRevLett.89.117901,Bal2012,RevModPhys.89.035002,cheng2023noisy}. The paradigm shift to the quantum level has offered new perspectives on quantum thermodynamics~\cite{RevModPhys.91.025001,Parrondo2015,Goold_2016,Popescu2014,PhysRevLett.129.230604}, which leads to the conceptualization and experimental scrutiny of various quantum models~\cite{Skrzypczyk_2011,PhysRevE.64.056130,PhysRevLett.125.070603,PhysRevLett.129.100603,PhysRevLett.122.070603,Correa2014,Steeneken2011,PhysRevLett.108.120602,Maslennikov2019,PhysRevLett.123.240601,PhysRevA.99.062103,Skrzypczyk2014,PhysRevLett.128.090602,Micadei2019}. At present, building an enhanced quantum refrigerator, characterized by heightened efficiency while remaining in compliance with the laws of thermodynamics, remains a paramount aspiration~\cite{PhysRevE.89.032115,10.1119/1.10023,PhysRevE.81.051129,doi:10.1126/science.1078955,geva1992quantum,Correa2013QuantumenhancedAR}.

Unlike classical refrigerators, quantum refrigeration demands a consideration of thermodynamic properties at the microscopic scale. A well-known example is the effective temperature of a single spin, which is defined by its polarization. This, in turn, introduces spin's ``cooling'' algorithms by enhancing its polarization through quantum operations~\cite{schulman1999,fernandez2004,Boykin2002,Rod2017}. The efficacy of this cooling method is intricately linked to the scale and correlation of the spin bath~\cite{PhysRevLett.119.050502}, rendering it a promising avenue for the development of robust quantum refrigeration systems. Drawing from a similar concept, a cooling model involving three qubits has been devised to explore the concept of the smallest conceivable refrigerator~\cite{PhysRevLett.105.130401}. This model incorporates the idea of self-containment, accentuating the refrigerator's operation without a reliance on external work sources~\cite{PhysRevLett.110.256801}.

In experimental studies, algorithmic cooling and related quantum refrigeration concepts have been demonstrated using nuclear magnetic resonance (NMR)~\cite{Baugh2005,PhysRevLett.100.140501} and nitrogen-vacancy centers in diamond~\cite{PhysRevLett.129.030601}. However, the realization of the self-contained refrigeration, operating without reliance on external resources, has remained elusive. In this work, we experimentally implement such a self-contained refrigerator model, comprising two cooling spins and one target spin. This configuration represents the smallest self-contained refrigerator achievable with two-level systems~\cite{PhysRevLett.105.130401}. Facilitated by the participation of three distinct spin-1/2 $^{13}$C nuclei in the same molecule, the entire cooling process requires no external input of work, owing to the meticulously designed gaps between energy levels. To understand the underlying mechanism of this refrigerator, we scrutinize the changes that occur within each spin during the cooling process and construct a refrigeration cycle. Furthermore, we conduct the working condition analysis and determine its achievable limiting temperature.

\emph{Model.}---
The concept of self-containment is introduced in exploring the minimal scale of thermal machines, employing three two-level quantum systems~\cite{PhysRevLett.105.130401}.  Let us start from the cooling process without the self-contained condition,  which involves two spins $q_1$ and $q_2$ at the same temperature $T_0$. They have zero ground-state energies, and excited-state energies at $E_1$ and $E_2$, where $E_1<E_2$. The population of the $i$-th spin ($i=1,2$) in the excited state is $ e^{-\beta E_i}/\mathcal{Z}_i$, where $\beta = 1/k_BT_0$ is the inverse temperature, $k_B$ is the Boltzmann constant, and $\mathcal{Z}_i = 1+e^{-\beta E_i}$ is the partition function.
At $T_0$, $q_1$ has a higher probability of being in the excited state compared to $q_2$ due to its smaller energy difference. If we apply a SWAP gate to exchange the states of $q_1$ and $q_2$, the temperature of $q_1$ becomes $T_0E_1/E_2$. Since $E_1<E_2$, the final temperature of $q_1$  is lower than its temperature $T_0$. As a result, $q_1$ undergoes cooling due to this population exchange.

In this particular scenario, the SWAP gate exchanges the two states, $\left|01\right> \leftrightarrow \left|10\right>$. It is evident that the SWAP gate necessitates an input of work due to the disparity in energy levels between these two states. Consequently, this cooling process for the two  spins cannot be considered self-contained. To achieve self-contained cooling, the energy levels being exchanged must be degenerate, ensuring that the operation consumes no work. This requirement, in turn, necessitates the introduction of a third spin, $q_3$, with an excited-state energy at $E_3$.

\begin{figure}[htbp]
    \centering
    \includegraphics[width=\columnwidth]{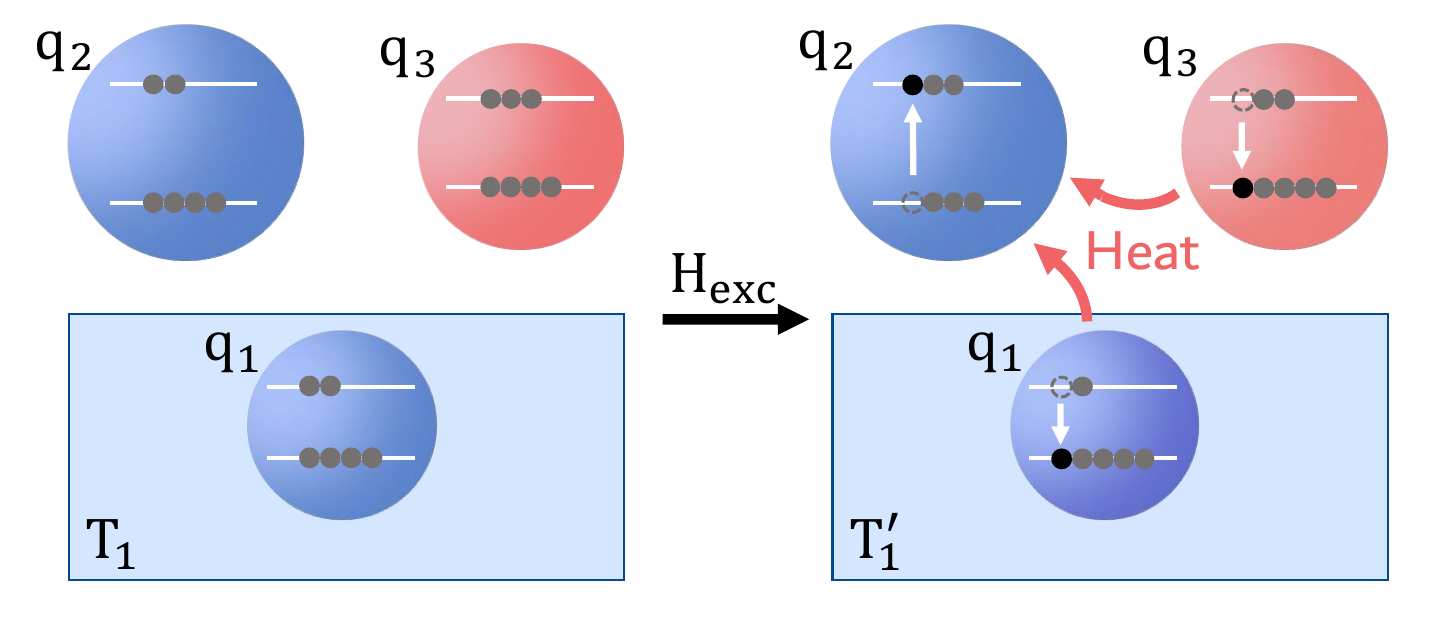}
\caption{Schematic representation of self-contained refrigeration designed to cool down the first spin, \(q_1\). The energy distribution among the three spins satisfies \(E_2=E_1+E_3\). The evolution under the Hamiltonian \(\mathcal{H}_{\text{exc}}\) exchanges the population by \(\left|010\right> \leftrightarrow \left|101\right>\). As a result, \(q_2\) extracts heat from the other two spins, leading to the cooling of the target spin \(q_1\) ($T'_1<T_1$).
}
    \label{fig:ac}
\end{figure}

When the energy relationship satisfies $E_2=E_1+E_3$, the joint states $\left|010\right>$ and $\left|101\right>$ become degenerate, meaning that the exchange of the two states requires no work input. Like the two-spin case, $q_1$ can be cooled down by this exchange, as depicted in Fig.~\ref{fig:ac}. It can be realized through the evolution under the interaction Hamiltonian $\mathcal{H}_{\text{exc}}=g(\left|010\right>\left<101\right|+\left|101\right>\left<010\right|)$, where $g$ represents the interaction strength~\cite{PhysRevLett.105.130401}. It is essential to note that such a Hamiltonian encompasses three-body interactions, which is not a natural term in real-world systems. Hence, it is imperative to employ control techniques that enable the realization of $\mathcal{H}_{\text{exc}}$ in a physical system while maintaining the self-containment, which will be discussed later in this Letter.

\emph{Experimental setup.}---
Now, we delve into the physical system designed to realize self-contained refrigeration. We employ a nuclear spin system, where three distinct two-level $^{13}$C nuclei are arranged in a chain-like molecular structure (crotonic acid dissolved in $d_6$-acetone)~\cite{PhysRevLett.129.070502,PhysRevLett.126.110502,supply}, to play the roles of the three spins in the refrigeration model. The experiment is conducted at room temperature using a Bruker 300 MHz NMR spectrometer. The internal Hamiltonian of the system can be written as $\mathcal{H}_{\text{NMR}}=-\sum_i\omega_i\sigma_z^i/2+\pi\sum_{i<j} J_{ij}\sigma_z^i\sigma_z^j/2,$ where $\sigma_{z}^i$ represents the Pauli matrix of the $i$-th spin, $\omega_i$ is the Larmor frequency, and $J_{ij}$ denotes the coupling strength between the $i$-th and $j$-th spins. Specific values for $\omega_i$ and $J_{ij}$ can be found in Supplemental Information~\cite{supply}. Additionally, we can apply transverse radio-frequency pulses to execute single-qubit rotations.

{Starting from the thermal equilibrium state dictated by the NMR system Hamiltonian $\mathcal{H}_{\text{NMR}}$, the first step involves creating the initial state $\rho_{0}$ for the refrigeration model~\cite{supply}.} In our experimental setup, we set the temperatures of three spins as $T_1=T_2=T_3/5$, where $T_1=2\delta/k_B$. The energy levels are defined as $E_1=\delta$, $E_2=3\delta$, and $E_3=2\delta$ to satisfy the self-contained condition $E_2=E_1+E_3$. Here, $\delta$ serves as an arbitrary energy unit, and we set $\delta=2$. Preparing $\rho_{0}$ under these specified parameters involves redistributing populations of $\rho_{\text{eq}}$ through the application of non-unitary gates. We utilize single-qubit rotations to redistribute the corresponding populations and apply 1 ms $z$-gradient pulses to eliminate unwanted coherence, as illustrated in Fig.~\ref{fig:work0}(a); see Supplemental Information for details~\cite{supply}.

\emph{Verification of self-containment.}---
The subsequent stage involves the evolution of the Hamiltonian $\mathcal{H}_{\text{exc}}=g(\left|010\right>\left<101\right|+\left|101\right>\left<010\right|)$. To facilitate its implementation in the experiment, we rewrite it as
\begin{equation}
\mathcal{H}_{\text{exc}}=\frac{g}{4}(\sigma^1_x\sigma^2_x\sigma^3_x+\sigma^1_x\sigma^2_y\sigma^3_y+\sigma^1_y\sigma^2_x\sigma^3_y+\sigma^1_y\sigma^2_y\sigma^3_x).
\label{Hexc}
\end{equation}
Remarkably, all terms in Eq.~(\ref{Hexc}) commute with each other, simplifying the simulation of this Hamiltonian and ensuring that the decomposition result is identical to that of the original Hamiltonian. Each three-body term can be further exactly decomposed into single- and two-qubit gates~\cite{PhysRevA.61.012302}, as illustrated in Fig.~\ref{fig:work0}(a), and a comprehensive pulse sequence is provided in Supplemental Information~\cite{supply}.

\begin{figure*}[htbp]
    \centering
    \includegraphics[width=2\columnwidth]{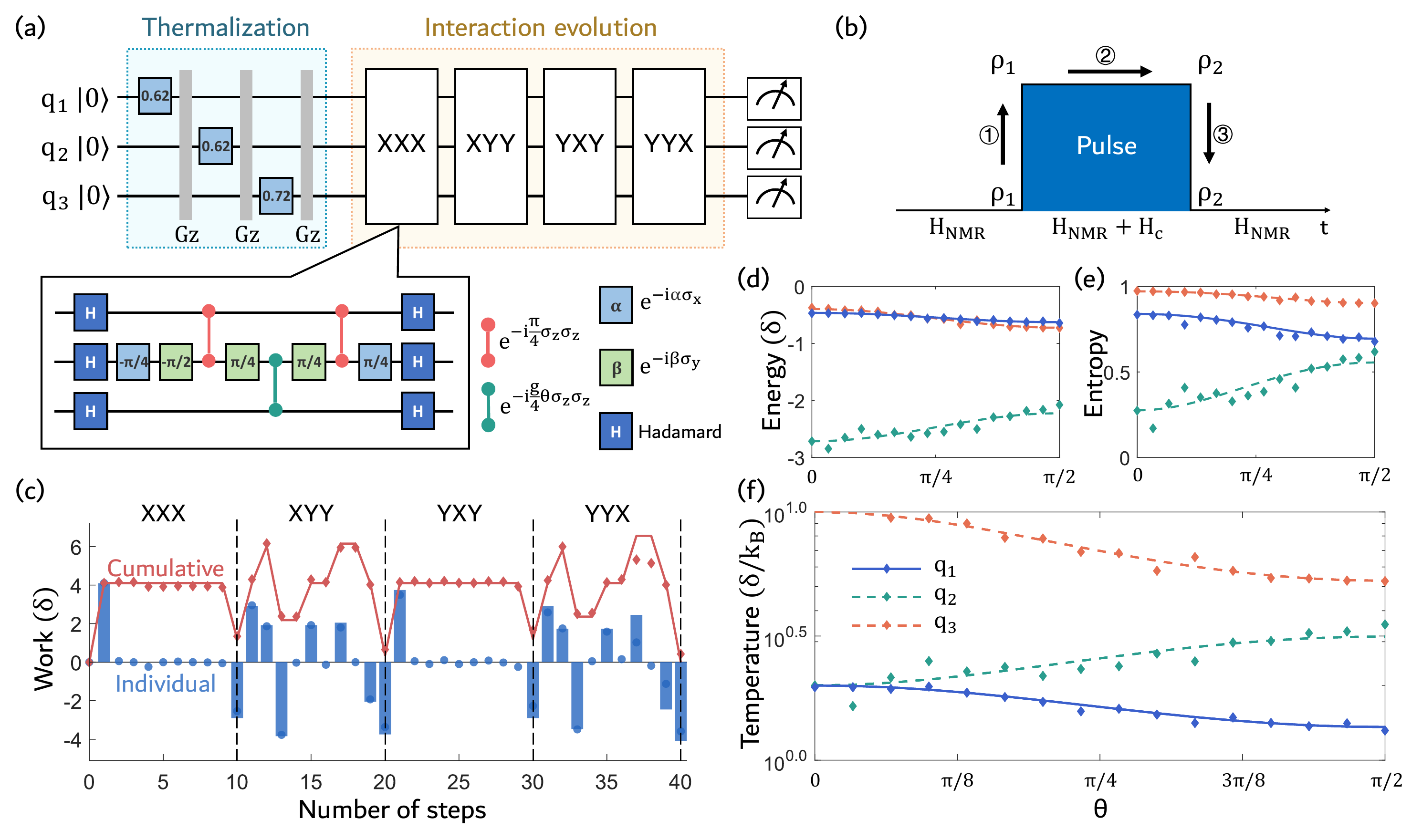}
\caption{(a) Quantum circuit for self-contained refrigeration. After preparing the initial thermal state $\rho_{0}$, the implementation of the $\mathcal{H}_{\text{exc}}$ evolution involves four three-body terms. Each term can be decomposed into ten steps comprising single- and two-qubit gates, as illustrated in the lower-left box. (b) Measurement of work and heat during the application of one pulse. The three steps correspond to $dW_1$, $dQ_1$, and $dW_2$, as defined in the main text. (c) Measured cumulative and individual work for the forty steps during the $\mathcal{H}_{\text{exc}}$ evolution. The total work sums up to zero when considering all individual work contributions, indicating that the refrigeration is self-contained. (d-f) Energy, entropy, and temperature of each spin during the $\mathcal{H}_{\text{exc}}$ evolution. The markers represent the experimental results. Entropy and energy transfer occur from $q_1$ and $q_3$ to $q_2$ over time, resulting in an increase in $q_2$'s temperature and a decrease in the temperatures of the others.}
    \label{fig:work0}
\end{figure*}

As $\left|010\right>$ and $\left|101\right>$ are degenerate, achieving their exchange during the evolution of $\mathcal{H}_{\text{exc}}$ does not require any input work. However, in our experimental setup, we employ the quantum simulation approach to realize this evolution, necessitating radio-frequency pulses and corresponding energy transfer.
To compute the energy transfer and verify the self-contained condition,
we measure work and heat separately in each experimental step. Here, work involves the energy transfer from the three-spin system to some external repository without changing entropy, while heat entails energy exchanges that involve entropy changes~\cite{PhysRevX.5.031044}. Specifically, work and heat can be expressed as $dW=\text{tr}(\rho d H)$ and $dQ=\text{tr}(Hd\rho)$, respectively~\cite{callen1991thermodynamics}.

Without loss of generality, we show the measurement of work and heat during the application of one pulse. As depicted in Fig.~\ref{fig:work0}(b),  the thermodynamic change can be summarized into three steps. Assuming that the state is $\rho_1$ before applying the pulse, and the current Hamiltonian is $\mathcal{H_\text{NMR}}$. The first step is to apply the pulse, which introduces an additional Hamiltonian term $\mathcal{H_\text{c}}$. As the state remains unchanged, only work, expressed as $dW_1=\text{tr}(\rho_1 \mathcal{H_\text{c}})$, is involved, and there is no heat transfer. Subsequently, in the second step, the state evolves under a new Hamiltonian, entailing only heat transfer $dQ_1=\text{tr}[(\rho_2-\rho_1)(\mathcal{H_\text{NMR}}+\mathcal{H_\text{c}})]$. Here, $\rho_2$ is the state after applying the pulse to $\rho_1$. In the third step, the pulse $\mathcal{H_\text{c}}$ vanishes, and the only work involved is given by $dW_2=-\text{tr}(\rho_2 \mathcal{H_\text{c}})$. Consequently, the application of a pulse can be viewed as a combination of the above work and heat transfer. {Moreover, it can be calculated that $dQ_1=0$ in the ideal case, so the net work equals to the change of internal energy of the three-spin system ~\cite{supply}.}

{In the experiment, we measure the change in internal energy to determine the work for each of the forty steps during the implementation of the $\mathcal{H}_{\text{exc}}$ evolution. These cumulative and individual changes in work during the implementation are illustrated in Fig.~\ref{fig:work0}(c). Notably, although some steps involve the consumption and storage of work, the total work amounts to zero upon accounting for all individual contributions. The result shows that the entire refrigeration process requires no input work, indicating its self-contained nature as we discussed in ~\cite{supply}.}


\emph{Performance.}---
Given the established setup, we have demonstrated that refrigeration occurs in a self-contained manner, which ultimately results in the cooling of the target spin $q_1$. However, the precise mechanisms governing this cooling process, as well as its performance under diverse conditions, remain unclear. Therefore, subsequent to preparing the thermal state for the entire system, we introduce a variable evolution time denoted as $\theta$ in $e^{-i\mathcal{H}_\text{exc} \theta}$ and monitor the state of each spin at various time points. In experiment, we measure the energy, entropy, and temperature of each spin at different $\theta$, and the obtained results are depicted in Figs.~\ref{fig:work0}(d-f).

\begin{figure}[htbp]
    \centering
    \includegraphics[width=\columnwidth]{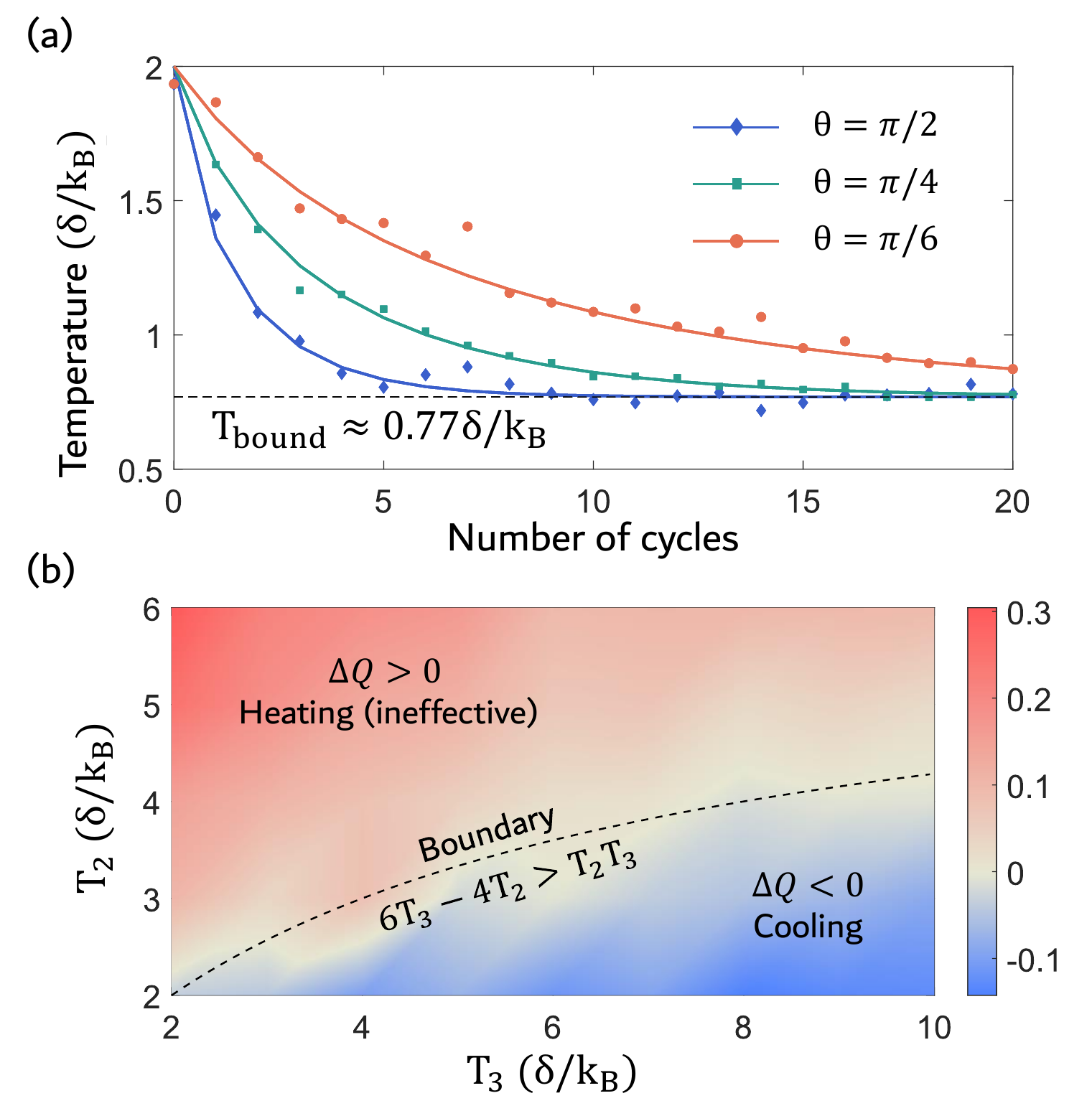}
\caption{(a) Temperature of the target spin $q_1$ for different numbers of refrigeration cycles at various evolution times $\theta$. As $\theta$ increases (up to $\pi/2$, indicating a complete population exchange), the cooling time becomes shorter. Eventually, all cases converge to the same temperature, approximately $T_\text{bound} \approx 0.77\delta/k_B$, regardless of the evolution time. (b) Experimental heat transfer as a function of the temperatures of $q_2$ and $q_3$. The blue region represents heat released from $q_1$, leading to its cooling, and the red region represents heat absorbed by $q_1$. The dashed line indicates the predicted boundary between the two cases.
}
    \label{fig:evowitht}
\end{figure}

As time progresses, the disorder of $q_1$ diminishes as entropy decreases, akin to the ``data compression" step in algorithmic cooling~\cite{Boykin2002}. Simultaneously, the energy of $q_1$ is transferred to the other spins, resulting in a reduction in its temperature. When considering all three spins collectively, despite the absence of work in this cooling process, it complies with the second law of thermodynamics. Energy transfer takes place, accompanied by a heat flow from the higher-temperature spin ($q_3$) to the lower-temperature spin ($q_2$), which has the potential to drive an ``engine". The refrigeration of $q_1$ is thus propelled by this engine  ~\cite{supply,Correa2014}. Consequently, spins $q_1$ and $q_3$ experience a decrease in temperature, while $q_2$ undergoes an increase. This thermodynamic perspective provides a clear picture of the refrigeration mechanism.

The refrigeration process can be extended into a cycle to achieve continuous cooling and attain and sustain the lowest temperatures. This cycle involves resetting $q_2$ and $q_3$  by bringing them into contact with their respective thermal environments and evolving the entire system under $\mathcal{H}_\text{exc}$ for the subsequent loop to further cool down the target spin $q_1$. {When resetting these spins through contact, the heat in $q_2$, transferred from $q_1$ during refrigeration, will be dissipated into the environment. $q_3$ will absorb heat from the environment to restore its original states, akin to the discussion in Supplemental Information~\cite{supply}.} Fig.~\ref{fig:evowitht}(a) illustrates the temperature as a function of the cycle number $n$ for various values of the evolution time $\theta$. As $n$ increases, the temperature of $q_1$  gradually decreases, indicating the effectiveness of the refrigeration cycle in continuously cooling down the target spin. Notably, with the evolution time $\theta$ increasing, the cooling efficiency improves, with  $q_1$ experiencing the quickest cooling when $\theta = \pi/2$. This observation is expected, considering that $\theta = \pi/2$ implies a more complete information exchange.

Of particular interest is the fact that, irrespective of the evolution time $\theta$ within one cycle, all temperatures tend to converge to the same bounded value $T_\text{bound}  \approx 0.77\delta/k_B$ after a finite number of cycles, as indicated by the dashed line in Fig.~\ref{fig:evowitht}(a). This convergence underscores the limitation of this refrigeration process under these specific conditions, and we will delve into its underlying mechanism in the subsequent discussion.

\emph{Working conditions.}---
The decrease in the temperature of $q_1$ results from the evolution under the Hamiltonian $\mathcal{H}_\text{exc}$, which exchanges the population of states: $\left|010\right> \leftrightarrow \left|101\right>$. To ensure a lower temperature, the polarization of $q_1$ needs to be enhanced, implying that the population of $q_1$'s excited state needs to decrease after this exchange. As only these two states are affected by this evolution, it is appropriate to exclusively consider them when analyzing the population change. For the initial thermal state, the populations of $\left|010\right>$ and $\left|101\right>$ are $P_{010}=e^{-\beta_2 E_2}/\mathcal{Z}_\text{total}$ and $P_{101}=e^{-\beta_1 E_1-\beta_3 E_3}/\mathcal{Z}_\text{total}$, respectively. Here, $\mathcal{Z}_\text{total} = \mathcal{Z}_\text{1}\mathcal{Z}_\text{2}\mathcal{Z}_\text{3}$ represents the total partition function. Obviously, $P_{010}$ corresponds to the probability for the ground state of $q_1$, while $P_{101}$ is for the excited state. To increase $q_1$'s polarization, $P_{101}>P_{010}$ is both a sufficient and necessary condition.

Consequently, the effective working condition for this refrigeration is given by
\begin{equation}\label{condition}
\frac{E_1}{T_1}+\frac{E_3}{T_3}<\frac{E_2}{T_2}.
\end{equation}
This condition bears a resemblance to the second law of thermodynamics for quantum refrigerators~\cite{PhysRevLett.108.070604}. Indeed, these concepts are interconnected through the specialized efficiency of this refrigerator~\cite{Correa2014}.

In the cooling cycle, spins $q_2$ and $q_3$ are reset after each cycle, and their temperatures remain initially equal at the cycle's outset. Therefore, these temperatures can be considered constant throughout the entire refrigeration process. The cooling limitations of this cycle can be determined using Eq.~(\ref{condition}) by fixing these two temperatures. We calculate that $T_1>T_\text{bound}$ is the condition for successful refrigeration. If the temperature of $q_1$  satisfies this requirement, it will be cooled down in the subsequent cycle. Otherwise, the refrigeration process fails. Thus, this limitation represents the lowest temperature that can be reached using this cooling cycle, consistent with the experimental results depicted in Fig.~\ref{fig:evowitht}(a). Importantly, this limitation is independent of the evolution time $\theta$ in each cycle, as the exchange is effective even in infinitesimally small time intervals. Once initiated, the temperature will continually decrease over subsequent cycles, eventually converging to this limitation.

This working condition is intricately linked to the temperatures of all three spins. The temperature of the target spin $q_1$  is particularly crucial as it establishes the cooling limitation over multiple cycles. Additionally, to comprehend how to maintain a system's low temperature, analyzing the refrigeration performance while keeping $q_1$'s temperature fixed becomes pivotal. To identify the proper conditions necessary for cooling the system at $2\delta/k_B$, we conduct experiments at varying temperatures $T_2$ and $T_3$. Maintaining the same energy differences as in the previous experiment and fixing $q_1$'s temperature at $T_1=2\delta/k_B$, we execute a refrigeration process with $T_3=[2\delta/k_B, 10\delta/k_B]$ and $T_2 = [2\delta/k_B, 6\delta/k_B]$. The heat transfer from $q_1$ is depicted in Fig.~\ref{fig:evowitht}(b). The cooling process operates effectively when heat is released from this spin. This outcome aligns with Eq.~(\ref{condition}). Consequently, we determine that the requirement for effective cooling is $6T_3-4T_2-T_2T_3>0$ (with temperature units set to $\delta/k_B=1$). Hence, we can establish the boundary where no heat transfer occurs at $q_1$ (dashed line in Fig.~\ref{fig:evowitht}(b)). The experimental results match well with the theoretical prediction, with the blue region signifying that refrigeration is valid under this setting and the red region indicating refrigeration failure.

\emph{Conclusion.}---
We demonstrate self-contained refrigeration using three nuclear spins through the NMR technique. In terms of dimensionality, this refrigerator is the smallest built using two-level systems~\cite{PhysRevLett.105.130401}. By measuring the energy transfer during the evolution of the three-body interaction, we show that no net external work is needed to drive the refrigeration process, implying its self-containment. Additionally, we analyze the performance and mechanism of this refrigeration approach by monitoring the changes in energy, entropy, and temperature throughout the evolution, and additionally obtain the working condition of this refrigeration. {Moreover, we delve into the potential implementation of this refrigeration cycle in quantum computation and its applicability within two solid-state spin systems~\cite{supply}. The insights gained from our analysis can not only be extended to other thermodynamic models but also prove beneficial for other physical scenarios requiring the analysis of energy transfer, such as algorithmic cooling~\cite{PhysRevLett.116.170501}.}

\begin{acknowledgments}
This work is supported by the National Key Research and Development Program of China (2019YFA0308100), National Natural Science Foundation of China (12104213, 12075110 and 12204230), Science, Technology and Innovation Commission of Shenzhen Municipality (JCYJ20200109140803865 and KQTD20190929173815000),  Guangdong Innovative and Entrepreneurial Research Team Program (2019ZT08C044), Guangdong Provincial Key Laboratory (2019B121203002), 2019QN01X298, and Guangdong Provincial Quantum Science Strategic Initiative (GDZX2303001 and GDZX2200001).
\end{acknowledgments}

\nocite{*}


%

\section{Appendix A: Algorithmic Cooling}
Algorithmic cooling is a computational technique used to transfer heat (or entropy) from certain qubits to either other qubits within the system or to the external environment, thereby achieving a cooling effect. Its foundation lies in the Basic Compression Subroutine (BCS), which is the simplest purification procedure in information theory. Let us commence by introducing BCS.

Consider a series of bits with bias $\epsilon_0$ ($\epsilon_0 > 0$) arranged in a line. Each bit is 0 with a probability of $P_0=(1+\epsilon_0)/2$ and 1 with a probability of $P_1=(1-\epsilon_0)/2$. Pairing them two by two, a controlled-NOT (CNOT) transformation is applied to determine whether they are the same or different. The resulting probabilities after the CNOT transformation are as follows:
\begin{equation}
\begin{aligned}
&P(0_c0_t\to0_c0_t) = \frac{(1+\epsilon_0)^2}{4},
\\
&P(0_c1_t\to0_c1_t) = \frac{1-\epsilon_0^2}{4},
\\
&P(1_c0_t\to1_c1_t) = \frac{1-\epsilon_0^2}{4},
\\
&P(1_c1_t\to1_c0_t) = \frac{(1-\epsilon_0)^2}{4}.
\end{aligned}
\end{equation}
After this transformation, retaining pairs that yield 0 on the target bit and swapping the control bit to the left, while for pairs that yield 1, leaving the control bit in its current place. Finally, discarding all target bits to the right, the bits on the left are purified to a bias of $\epsilon_1=2\epsilon_0/(1+\epsilon_0^2)$ from $\epsilon_0$ with a probability of $\epsilon_0/(2\epsilon_1)$.

The BCS operation is denoted as $\mathcal{B}$. Furthermore, in a physical system, we can establish a thermal contact with a heat bath to achieve the thermal state with the bias of $\epsilon_0$. This process can be denoted as $\mathcal{M}_0$. Algorithmic cooling is constructed by repeating these two operations. Starting with a long string of bits, the thermal contact operation is applied to the first $m$ bits to reach a bias of $\epsilon_0$. Subsequently, a BCS operation $\mathcal{B}$ is applied to the first $m$ bits, resulting in purified bits on the left with an expected count of $m\epsilon_0/(2\epsilon_1)$. After this process, $m/2$ disused bits with increased entropy remain on the right. This constitutes a small cycle of algorithmic cooling.

For the next cycle, a different set of $m$ bits is chosen from $1+m/2$ to $(3m)/2$ to undergo these two operations. After repeating this cycle $l$ times, the anticipated quantity of bits purified to $\epsilon_1$ can be formulated as:
\begin{equation}
\left< \mathcal{L}^l\right>=\frac{lm(1+\epsilon_0^2)}{4}.
\end{equation}
For $l \ge 4$, it is guaranteed that $\left< \mathcal{L}^l\right> \ge m$ when $m$ is sufficiently large. Then, a cut operation, denoted as $\ell_j$, is defined to preserve only the initial purified $m$ bits.

At the end of the previous cycle, we obtain $m$ bits with a bias of $\epsilon_1$, resembling a thermal contact with another heat bath at a lower temperature. Thus, this cycle can be regarded as another thermal contact operation $\mathcal{M}_1$, representing the first iteration of algorithmic cooling. Generally, the expression for the $j$-th algorithmic cooling is given by:
\begin{equation}
M_j=\ell_j\underbrace{\mathcal{B}\mathcal{M}_{j-1}\dots\mathcal{B}\mathcal{M}_{j-1}}_{l\ \text{times}}.
\end{equation}
After the $j$-th algorithmic cooling, the bias of the first $m$ bits follows the iteration relation:
\begin{equation}
\epsilon_j=2\epsilon_{j-1}/(1+\epsilon_{j-1}^2).
\end{equation}
Therefore, for any choice of the destination temperature corresponding to a bias $\epsilon_{\text{des}}$, one can calculate the minimal $j_m$ such that $\epsilon_{j_m}\ge\epsilon_{\text{des}}$. Applying the entire $j_m$-th algorithmic cooling yields systems at the desired temperature.

\section{Appendix B: Details for Self-Contained Refrigeration}
The concept of self-contained systems was introduced in the exploration of the minimal scale of thermal machines. To simplify the analysis, we begin with a cooling process involving two distinct spins represented by the Hamiltonians $\mathcal{H}_1$ and $\mathcal{H}_2$ respectively, where $\mathcal{H}_i=E_i\left|1\right>\left<1\right|$, $\left|1\right>$ denotes the excited states at energies $E_i$, and $\left|0\right>$ represents the ground state at zero energy. The energy of the excited state of the second spin is higher than that of the first one ($E_1 < E_2$). The thermal state of the systems is determined by temperature and energy, given by the formula:
\begin{equation}
\tau=e^{-\beta \mathcal{H}}/\mathcal{Z},
\end{equation}
where $\mathcal{Z}=\mathrm{Tr}(e^{-\beta \mathcal{H}})$ is the partition function, and $\beta=1/k_B T$ represents the inverse temperature. Therefore, the population of the excited states can be calculated as $P^e_i={e^{-\beta E_i}}/{(e^{-\beta E_i}+1)}$.

If these two spins are initially at the same temperature $T_0$, the first spin has a higher probability of being in the excited state compared to the second one ($P^e_1 > P^e_2$). If a SWAP operation is performed to exchange these spins, the distribution of the second spin is transferred to the first one. The population of the excited state for the latter spin decreases, and can be expressed as:
\begin{equation}
P^e_1=\frac{e^{-\beta_0 E_2}}{e^{-\beta_0 E_2}+1}=\frac{e^{-\beta' E_1}}{e^{-\beta' E_1}+1}.
\end{equation}
By comparing the population expressions, we can determine the change in temperature for the first spin, denoted as $T'=T_0E_1/E_2$, which is lower than the initial temperature. Consequently, the first spin is cooled down due to this exchange.

However, for the cooling process to be self-contained, the SWAP operation must be devoid of work. This is not achievable with the two-level system described above, necessitating the study of a three-spin system.

Consider three spins with the energy difference $E_i$ as the excited energy and zero energy for the ground state. Refrigeration is achieved by exchanging states $\left|010\right>$ and $\left|101\right>$. The condition ensuring the degeneracy of these two levels is:
\begin{equation}
E_2=E_1+E_3.
\end{equation}
However, at the given inverse temperature $\beta_l$, the population of these two levels is equal:
\begin{equation}
P_{101}=\frac{e^{-\beta_l E_1}e^{-\beta_l E_3}}{Z_s}=\frac{e^{-\beta_l E_2}}{Z_s}=P_{010}.
\end{equation}
Exchanging these two states changes nothing. Differently, this will be changed by assigning a higher temperature to the third spin. The inverse temperature will be $\beta_h < \beta_l$ for this new condition. The population of state $\left|101\right>$, $P_{101}$, increases and exceeds $P_{010}$. As a result, $P_{101}$ decreases with this exchange operation, and the population of the excited state of the first spin will decrease. This means the temperature of the first spin gets lower from $T_l$. Thus, self-contained refrigeration is achieved.

\section{Appendix C: Experimental Details}

\begin{figure}[htbp]
\centering
\includegraphics[width=1\linewidth]{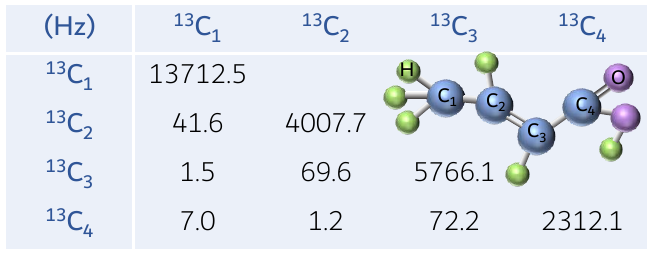}
\caption{Molecular structure and parameters of the four-qubit NMR quantum processor. The Larmor frequencies are listed on the diagonal, while the coupling strengths are provided in the off-diagonal positions.}
\label{molecular}
\end{figure}

Our experiments were conducted utilizing a nuclear magnetic resonance (NMR) quantum processor with $^{13}$C-labeled crotonic acid dissolved in deuterated acetone (d6-acetone).  
The four coupled $^{13}$C spins in the molecular structure serve as qubits, with internal Hamiltonian in the form of:
\begin{equation}
\mathcal{H}_{\text{NMR}} = -\sum^4_{i=1}\frac{\omega_i}{2}\sigma^i_z + \pi\sum^4_{i<j}\frac{J_{ij}}{2}\sigma^i_z\sigma^j_z,
\end{equation}
where $\sigma_z^i$ denoteds the z-Pauli matrix for the $i$-th spin, $\omega_i$ is the Larmor frequency of the $i$-th spin, and $J_{ij}$ denotes the coupling strength between the $i$-th and $j$-th spins. The molecular structure and the Hamiltonian parameters are illustrated in Figure~\ref{molecular}. In our experiment, we employed three nuclei C$_{2-4}$:  C$_2$ represents the spin requiring cooling, while C$_3$ and C$_4$ are spins in contact with two separate heat reservoirs at low and high temperatures, respectively.

Control of this system is achieved by applying radio-frequency (rf) pulse, as described by the Hamiltonian:
\begin{equation}
\begin{aligned}
\mathcal{H}_{\text{rf}} = -\sum_i\gamma_iB_i[\cos{((\omega_{\text{rf}}-\omega_i)t+\phi)} \sigma^i_x \\
+\sin{((\omega_{\text{rf}}-\omega_i)t+\phi)} \sigma^i_y],
\end{aligned}
\end{equation}
where $B_i$, $\phi$, and $\omega_{\text{rf}}$ denote the amplitude, phase, and frequency of the pulse, respectively. Arbitrary single spin operations are realized through a series of pulses with varying amplitudes and phases.

\begin{figure*}[htbp]
\centering
\includegraphics[width=1\linewidth]{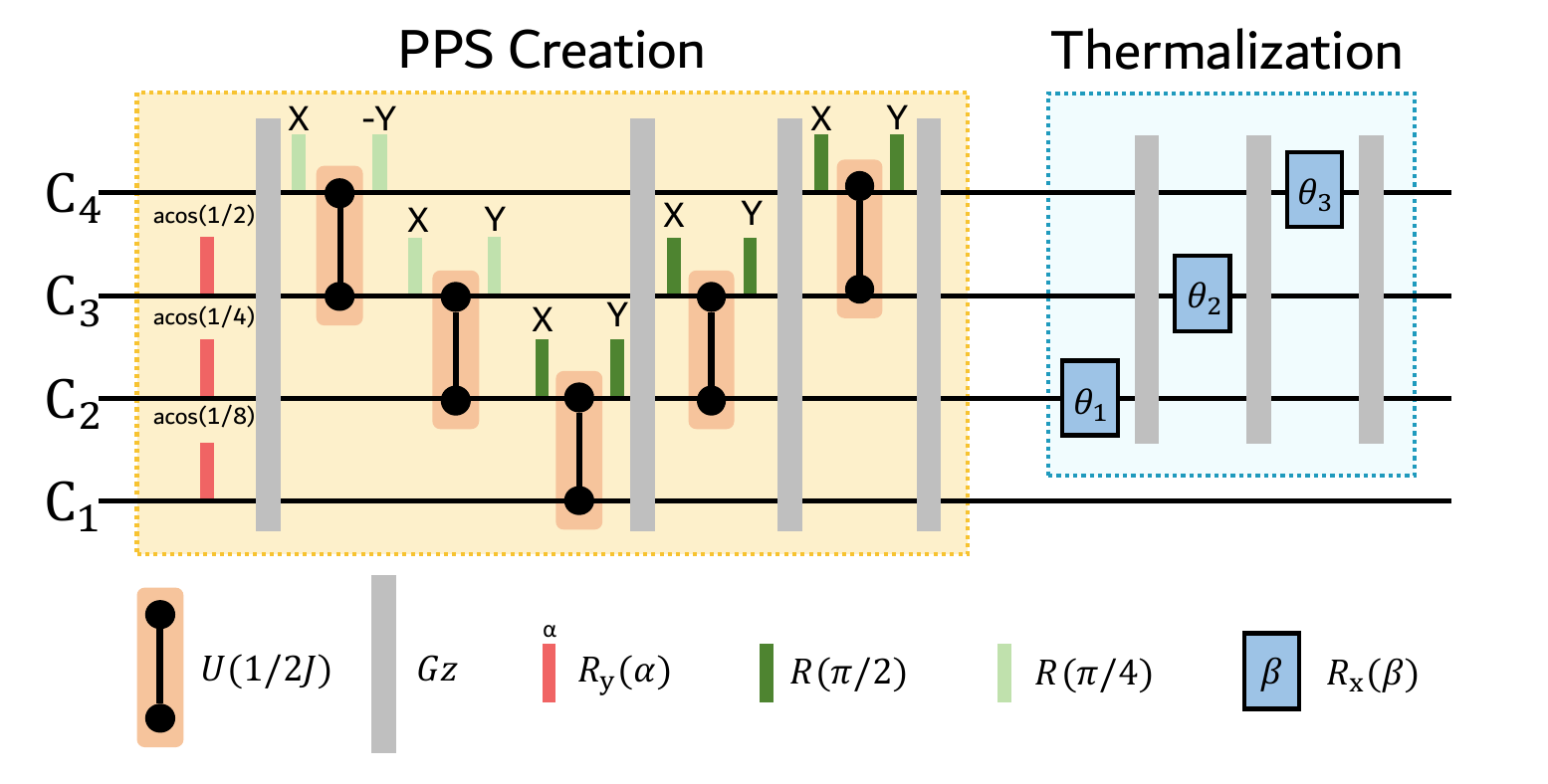}
\caption{Circuit to prepare the initial thermal state. The operations in the yellow box are employed to prepare PPS using the spatial average method, while the operations in the blue box are utilized to prepare a thermal state at a specific temperatures from PPS. }
\label{circuit}
\end{figure*}

The experiments were performed at room temperature.  The corresponding thermal state of this spin system is described as follows:
\begin{equation}
\rho_{\text{eq}} = \frac{I}{2^4} + \epsilon\sum^4_{i=1}\sigma_z^i,
\end{equation}
where $I$ is the $16\times 16$ identity matrix, and the polarization $\epsilon$ is on the order of $10^{-5}$. We utilized the spatial average method,   utilizing a pulse sequence illustrated within the yellow rectangle of Figure~\ref{circuit}, to initialize the NMR system to a pseudo-pure state (PPS):
\begin{equation}
\rho_{\text{PPS}} = \frac{1-\epsilon}{16}I + \epsilon\left|0000\right>\left<0000\right|.
\end{equation}
The identity matrix term is invariant under any unitary operation and does not produce a signal when detected in NMR. Hence, the quantum system can be treated as a pure state $\left|0000\right>\left<0000\right|$, up to a scale factor. To prepare the thermal state at any specific temperature $\rho_0=\rho_{1}\otimes\rho_{2}\otimes\rho_{3}$ (where $\rho_i$ represents the thermal state at temperature $T_i$), a thermalization process is applied to the PPS. This thermalization process for each spin consists of two steps: the first is a $\theta=\arccos(\sqrt{P^g_{i}})$ rotation along the $x$-axis to adjust the populations in the ground and excited states, where $P^g_{i}$represents the population of the ground state for the thermal state at temperature $T_i$. Then, gradient pulses are used to eliminate the coherence, resulting in a thermal state. The entire process to prepare $\rho_0$ is illustrated in Figure~\ref{circuit}.

During the measurement, the signal from each spin is typically split into eight peaks, arising from couplings between different nuclei. These peaks consist of both real and imaginary parts, with the real part encoding the expectation values of the observable spin's Pauli matrix $\sigma_x$ and the imaginary part encoding the expectation values of $\sigma_y$. This enables the measurement of expectation values of single-quantum coherence operators involving $\sigma_x$ or $\sigma_y$ in the target spin, and $\sigma_z$ or $I$ in the other spins for one measurement. By applying readout pulses, other operators can also be measured by changing their basis, indicating that full quantum state tomography is available.

\section{Appendix D: Decomposition of Evolution of the three-body Hamiltonian}
The Hamiltonian under consideration is expressed as the sum of Pauli operators:
\begin{equation}
\begin{aligned}
\mathcal{H}_i &= g\left(|010\rangle\langle101| + |101\rangle\langle010|\right) \\
&= \frac{g}{4}\left(\sigma_x^1\sigma_x^2\sigma_x^3 + \sigma_x^1\sigma_y^2\sigma_y^3 + \sigma_y^1\sigma_x^2\sigma_y^3 + \sigma_y^1\sigma_y^2\sigma_x^3\right).
\end{aligned}
\end{equation}
For the four terms on the right commute with each other, the evolution $e^{-i\mathcal{H}_i\theta}$ can be decomposed into four evolutions, irrespective of the order:
\begin{equation}\label{Eq14}
e^{-i\mathcal{H}_i\theta} = e^{-i\frac{g}{4}\theta\sigma_x^1\sigma_x^2\sigma_x^3}e^{-i\frac{g}{4}\theta\sigma_x^1\sigma_y^2\sigma_y^3}e^{-i\frac{g}{4}\theta\sigma_y^1\sigma_x^2\sigma_y^3}e^{-i\frac{g}{4}\theta\sigma_y^1\sigma_y^2\sigma_x^3}.
\end{equation}

However, these products of Pauli operators still correspond to three-body interactions, which are not typical Hamiltonians in spin systems. To resolve this issue, it is necessary to decompose them further into one- or two-body Hamiltonians.
Specifically, a three-body interaction involving Pauli $\sigma_z$ can be achieved through eight evolutions under different Hamiltonians:
\begin{equation}
\begin{aligned}
e^{-i\frac{\pi}{2}J_{123}\theta\sigma^1_z\sigma^2_z\sigma^3_z} &= e^{-i\frac{\pi}{4}\sigma^2_x}e^{-i\frac{\pi}{4}\sigma^1_z\sigma^2_z}e^{-i\frac{\pi}{4}\sigma^2_y}e^{-i\frac{\pi}{2}J_{123}\theta\sigma^2_z\sigma^3_z}\\
&\quad \times e^{-i\frac{\pi}{4}\sigma^2_y}e^{-i\frac{\pi}{4}\sigma^1_z\sigma^2_z}e^{i\frac{\pi}{2}\sigma^2_y}e^{i\frac{\pi}{4}\sigma^2_x}.
\end{aligned}
\end{equation}
The operators $\sigma_x$ and $\sigma_y$ are related to $\sigma_z$ through $H\sigma_zH$ and $H_y\sigma_zH_y$, where $H$ and $H_y$ are basis-changing matrices defined in the computational basis as:
\begin{equation}
\begin{aligned}
&H = \frac{1}{\sqrt{2}}
\begin{bmatrix}
1 & 1\\
1 & -1
\end{bmatrix},
\\
&H_y = \frac{1}{\sqrt{2}}
\begin{bmatrix}
1 & -i\\
i & -1
\end{bmatrix}.
\end{aligned}
\end{equation}
All four three-body Pauli interaction Hamiltonians in Eq.~(\ref{Eq14}) can be simulated within an NMR spectrometer. Each one-body Pauli interaction can be decomposed into ten steps, taking $e^{-i\frac{g}{4}\sigma_x\sigma_x\sigma_x\theta}$ as an example:
\begin{equation}
\begin{aligned}
e^{-i\frac{g}{4}\theta\sigma^1_x\sigma^2_x\sigma^3_x} &= H^{123}e^{-i\frac{\pi}{4}\sigma^2_x}e^{-i\frac{\pi}{4}\sigma^1_z\sigma^2_z}e^{-i\frac{\pi}{4}\sigma^2_y}e^{-i\frac{g}{4}\theta\sigma^2_z\sigma^3_z}\\
&\quad \times e^{-i\frac{\pi}{4}\sigma^2_y}e^{-i\frac{\pi}{4}\sigma^1_z\sigma^2_z}e^{i\frac{\pi}{2}\sigma^2_y}e^{i\frac{\pi}{4}\sigma^2_x}H^{123}.
\end{aligned}
\end{equation}
Here, $H^{123}$ represents the Hadamard gates applying to the three spins individually. Thus, we can achieve the evolution $e^{-i\mathcal{H}_i\theta}$ through forty steps of one- and two-body operations.

\section{Appendix E: Conservation of Heat in Quantum Evolution}

Ensuring heat conservation during the evolution process is imperative for the reliability of our results. It ensures that the heat remains invariant throughout the simulation, and any changes in energy are solely attributed to the work done when altering the system's environment to perform operations. This assertion can be substantiated by analyzing the energy changes throughout the process.

In accordance with the first law of thermodynamics, the internal energy of a system comprises both heat and work. In the realm of quantum thermodynamics, work is influenced by the Hamiltonian, while heat is influenced by the state, expressed as $dE=dW+dQ=\rho d\mathcal{H}+\mathcal{H} d\rho$. As the evolution progresses under a specific Hamiltonian, $\mathcal{H}$ remains constant. Consequently, the change in heat is equivalent to the change in internal energy.
\begin{equation}
\begin{aligned}
dQ=dE&=\mathrm{Tr}(\rho' \mathcal{H})-\mathrm{Tr}(\rho \mathcal{H})\\
&=\mathrm{Tr}(e^{-i\mathcal{H}\theta} \rho e^{i\mathcal{H}t} \mathcal{H})-\mathrm{Tr}(\rho \mathcal{H})=0.
\end{aligned}
\end{equation}
This implies that both heat and internal energy are conserved during the evolution. The change in internal energy for an operation is identical to the work $dE=\sum dW$.

\section{Appendix F: Self-containment in our simulation of the evolution}
In this section, we will elucidate the self-contained nature from two perspectives. Firstly, through a numerical analysis, we will discuss why the total input work during our experimental pulse sequence is zero. Secondly, we will construct a comprehensive physical picture analogous to our experimental implementation to illustrate the meaning of positive or negative work.

\begin{figure*}[htbp]
    \centering
    \includegraphics[width=\linewidth]{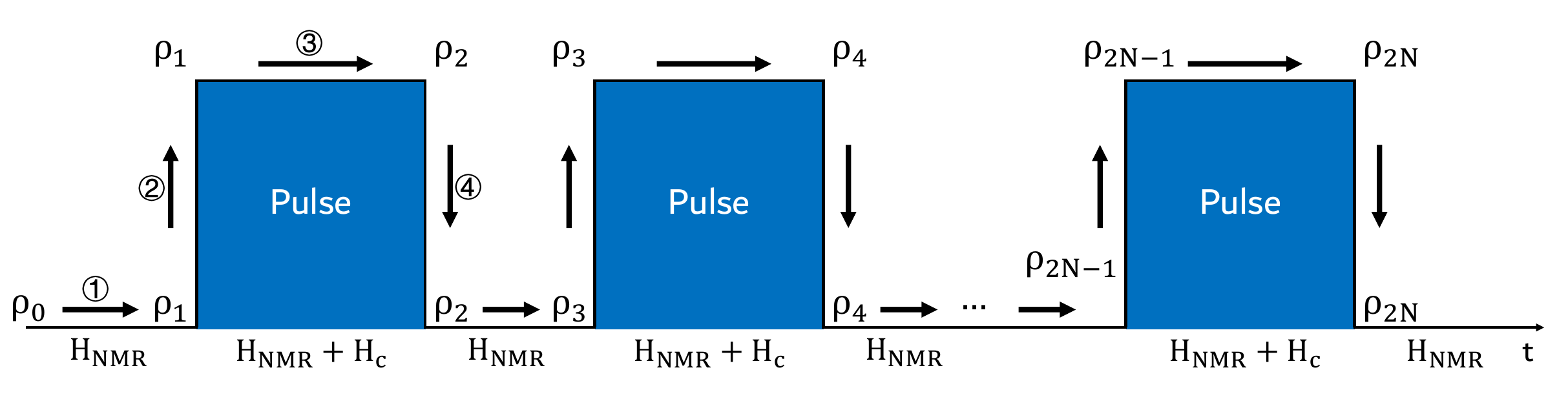}
    \caption{{Sequence of pulses to realize the refrigeration, detailing the four steps involved in applying the first pulse, numbered from \textcircled{1} to \textcircled{4}.}}\label{Rf4}
\end{figure*}

For the first, we begin with the analysis of the energy changes in the real experiment. In our NMR experiments, the refrigeration is achieved by applying a sequence of pulses, as depicted in Figure~\ref{Rf4}. This is also the full version of the single-pulse case [Fig. 2(b) in the main text]. Each application of a pulse consists of four distinct phases, characterized by varying parameters such as the duration of intervals and pulse parameters, which are tuned to guide the system's evolution towards the desired operation. These phases include two intervals of free evolution and two intervals where the Hamiltonian is altered, sequentially labeled from \textcircled{1} to \textcircled{4}.

During the free evolution \textcircled{1}, there is no applied pulse, i.e., no external control field. Consequently, the system's state undergoes the evolution dictated by the intrinsic Hamiltonian of the molecule, denoted as $\mathcal{H}_{\text{NMR}}$. Since $\mathcal{H}_{\text{NMR}}$ remains constant, there is no work done during the free evolution, as defined by $dW=\mathrm{tr}(\rho d\mathcal{H})$. Heat exchange during this phase is expressed as $dQ=\mathrm{tr}(\mathcal{H} d\rho)$. Given the changes in the state, the heat transfer in phase \textcircled{1} can be expressed as
\begin{equation}
\begin{aligned}
    dQ_1&=\mathrm{tr}\left[ \mathcal{H}_\mathrm{NMR}(\rho_1-\rho_0) \right]\\
&=\mathrm{tr}(\mathcal{H}_\mathrm{NMR}e^{-i\mathcal{H}_\mathrm{NMR}t_1}\rho_0 e^{i\mathcal{H}_\mathrm{NMR}t_1})-\mathrm{tr}(\mathcal{H}_\mathrm{NMR}\rho_0).
\end{aligned}
\end{equation}
Here, $\rho_0$ is the initial state, $t_1$ is free evolution time, and $\rho_1 = e^{-i\mathcal{H}_\mathrm{NMR}t_1}\rho_0 e^{i\mathcal{H}_\mathrm{NMR}t_1}$ is the state at the end of this phase. The outcome of this equation is zero, i.e., $dQ_1=0$, consistent with the commutation relation $[\mathcal{H}, e^{-i\mathcal{H}t}] = 0$.

The subsequent phase \textcircled{2} involves changes in the Hamiltonian, executed so rapidly that the state's variation is negligible, precluding heat exchange. The introduction of an external control field adds a term $\mathcal{H}_c$ to the Hamiltonian, resulting in work performed on the system as follows
\begin{equation}
    dW_1=\mathrm{tr}\left\{ \left[(\mathcal{H}_\mathrm{NMR}+\mathcal{H}_c)-\mathcal{H}_\mathrm{NMR}\right]\rho_1\right\}=\mathrm{tr}(\mathcal{H}_c\rho_1).
\end{equation}

The third phase \textcircled{3} mirrors the first in terms of free evolution, albeit with the Hamiltonian now including the control field term. This Hamiltonian also remains time-independent, thus no heat transfer occurs, as demonstrated by
\begin{equation}
    dQ_2=\mathrm{tr}\left[ (\mathcal{H}_\mathrm{NMR} +\mathcal{H}_c )(\rho_2-\rho_1)\right]=0.
\end{equation}

In the final phase \textcircled{4}, the control field is turned off, stabilizing the state $\rho_2$. Similar to the second phase, this leads to no heat exchange. The work performed on the system is given by
\begin{equation}
    dW_2=\mathrm{tr}\left\{ \left[\mathcal{H}_\mathrm{NMR} -(\mathcal{H}_\mathrm{NMR} +\mathcal{H}_c)\right]\rho_2\right\}= -\mathrm{tr}(\mathcal{H}_c\rho_2).
\end{equation}

Considering all four phases, the net change in internal energy for the application of a single pulse is
\begin{equation}
    dE_1=dQ_1+dW_1+dQ_2+dW_2=\mathrm{tr}\left[\mathcal{H}_c(\rho_1-\rho_2)\right].
\end{equation}
This reflects the cumulative work done on the system across all four phases. It is worth noting that, in the ideal case (the absence of experimental errors), the heat exchange is zero during the pulse application. Therefore, the net change in internal energy equals the total work consumption. Hence, summing the energy changes across the entire pulse sequence yields
\begin{equation}
    \Delta E=\Delta W+\Delta Q= \Delta W.
\end{equation}

Therefore, if $\Delta E=0$ holds for this task, we can infer that the total work consumption $\Delta W=0$ without accounting for experimental errors. To compute $\Delta E$, let us revisit the initial task, which involves the realization of self-contained refrigeration through the evolution of $\mathcal{H}_{\text{exc}}=g(\left|010\right>\left<101\right|+\left|101\right>\left<010\right|)$. The system starts from $\rho_0=e^{-\beta_1\sigma^1_z}\otimes e^{-\beta_2\sigma^2_z}\otimes e^{-\beta_3\sigma^3_z}/Z_1Z_2Z_3$, composed by the thermal states of three Hamiltonian $\mathcal{H}_i=E_i\sigma^i_z$ at their inverse temperature $\beta_i\ (i=1,2,3)$ with the constraint of $E_2=E_1+E_3$. By evolving under $\mathcal{H}_{\text{exc}}$, the population of levels $\left|010\right>$ and $\left|101\right>$ are exchanged. For the degeneration of these two levels, the net change in internal energy during this task is $\Delta E  = 0$. In our experiment, we employ the pulse sequence in Figure~\ref{Rf4} to quantum simulate the above task, dividing the evolution governed by the three-body Hamiltonian $\mathcal{H}_{\text{exc}}$ into forty distinct steps. This is an exact quantum simulation without any approximation in theory (see Appendix D in the Supplemental Information). Therefore, the experimental net change in internal energy should be the same as that of the initial task, i.e., $\Delta E = 0$ in the ideal case.

Above is the analysis of the self-contained condition (no net input work). In real experiments, we measure the energy changes at each of the forty steps and show that the cumulative work in experiment is very close to zero [see Fig. 2(c) in the main text], indicating the self-contained nature of this refrigeration process.

\begin{figure}[htbp]
    \centering
    \includegraphics[width=1\linewidth]{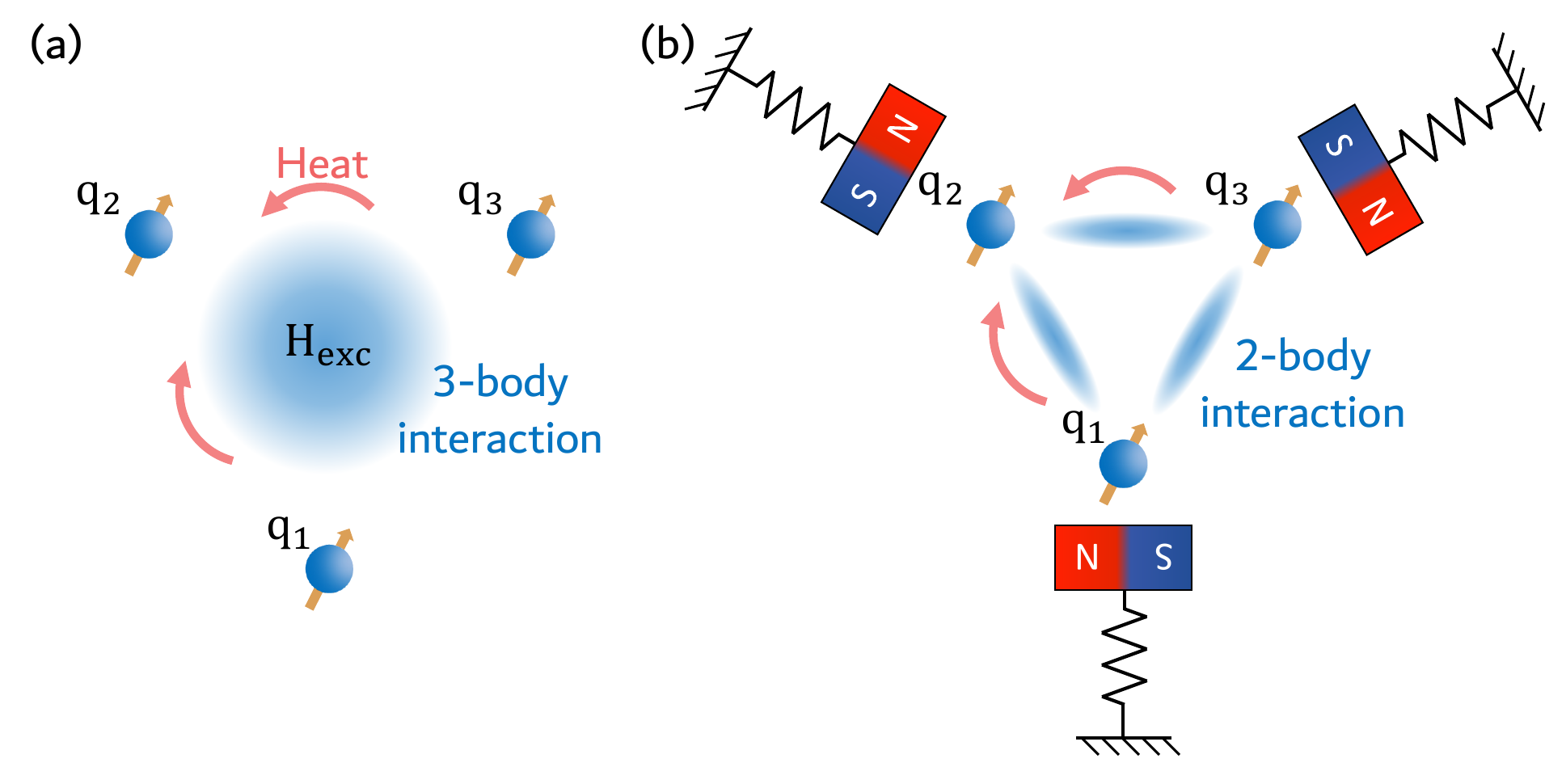}
    \caption{{Comprehensive physical models of self-contained refrigeration. (a) Original model featuring the three-body Hamiltonian $\mathcal{H}{\text{exc}}$. Refrigeration in this model requires no input work. (b) In our experiment, we combine pulses and natural evolutions of the system to simulate $\mathcal{H}_{\text{exc}}$. The application of a pulse can be visualized as small magnets attached to springs for the three spins. The work can be either positive or negative, depending on the direction of the work's flow. When the entire sequence concludes, the magnets and springs return to their original states, signifying that the total work is zero throughout the process.}}\label{Rf5}
\end{figure}

Now, we would like to construct a comprehensive physical picture analogous to our experimental implementation to illustrate the meaning of positive or negative work, as depicted in Figure~\ref{Rf5}. The core of the original model, as shown in Figure~\ref{Rf5}(a), is to evolve the three spins under the three-body Hamiltonian $\mathcal{H}_{\text{exc}}=g(\left|010\right>\left<101\right|+\left|101\right>\left<010\right|)$. Due to the degeneracy between $\left|010\right>$ and $\left|101\right>$, the refrigeration requires no input work from external resources.

In our experiment, we employ a sequence of pulses in conjunction with the natural evolution of $\mathcal{H}_\mathrm{NMR}$ containing two-body interactions to simulate $\mathcal{H}_{\text{exc}}$. During the application of each pulse, we activate a radio-frequency external magnetic field for a specific duration and subsequently deactivate it. In a comprehensive representation [see Figure~\ref{Rf5}(b)], this external magnetic field can be conceptualized as small magnets attached to springs. Additionally, the magnetic field exerts distinct influences on different spins due to their varying resonant frequencies.

As the magnet moves away from the spins, it imparts work to them, resulting in positive work values. Conversely, the spins impart work to the magnet, and this energy is conserved as potential energy in the spring when the magnet approaches these spins. In this scenario, the work value for spins is negative. Therefore, as the system undergoes the entire pulse sequence, individual work values can be either positive or negative for different pulses. When the sequence concludes, the magnets and springs return to their original states, signifying that the total work is zero throughout the entire process.

\section{Appendix G: Constraints Imposed by Thermodynamic Laws}
While a self-contained refrigerator can achieve cooling without the need for external work, it is important to note that heat transfer remains a crucial aspect of this operation. Even in this seemingly self-sustaining cooling process, it is subject to the constraints imposed by the laws of thermodynamics. In the following section, we will delve into the intricacies of this cooling process from a classical standpoint and elucidate how these fundamental thermodynamic laws both govern and impose limitations on the refrigerator's operation.

No classical machine can achieve cooling solely through heat flow. The self-contained quantum cooling demonstrated here is, in fact, a combination of a heat engine and a refrigerator. The transfer of heat from a hotter qubit to a cooler one, with a portion of the heat converting into work, constitutes a thermal machine. This work can then be employed to facilitate the cooling process. The coefficient of performance (COP) of this quantum refrigerator is defined as the ratio between the heat transfer from the cold bath in refrigeration and the heat flow from the hotter qubit , denoted as $\eta=Q_c/Q_h$. 
The maximum efficiencies permitted by thermodynamic laws for both the refrigerator and engine are the Carnot efficiency. Thus, the limitation of the COP for this combined machine can be expressed as the product of the Carnot engine and refrigerator efficiency, as given by:
\begin{equation}
\eta_m = \frac{(T_3 - T_2)T_1}{T_3(T_2 - T_1)},
\end{equation}
where $T_1$, $T_2$, and $T_3$ are temperatures of the three spins, satisfying $T_1 \le T_2 < T_3$. 

In self-contained refrigeration, the cooling process is governed by the Hamiltonian $\mathcal{H}_i=g(|010\rangle\langle101| + |101\rangle\langle010|)$. Here, only the populations of states $|101\rangle$ and $|010\rangle$ undergo changes. For effective cooling, the former decreases, while the latter increases by the same amount $\Delta P$. When tracing out the other system to obtain the state of a single spin, the changes in density matrices  of the first and third spins are identical:
\begin{equation}
\Delta\rho_1 = \Delta\rho_3 =
\begin{bmatrix}
\Delta P & 0\\
0 & -\Delta P
\end{bmatrix}.
\end{equation}
The COP of the self-contained refrigerator can be expressed by the ratio between the heats of these spins as:
\begin{equation}
\eta=\frac{\Delta Q_1}{\Delta Q_3} = \frac{\mathrm{Tr}(\mathcal{H}_1 \Delta\rho_1)}{\mathrm{Tr}(\mathcal{H}_3 \Delta\rho_3)} = \frac{E_1}{E_3},
\end{equation}
which is independent of temperature. 

Will the COP of the self-contained refrigerator $\eta$ exceed the limit imposed by Carnot's theorem $\eta_{m}$?
 It may initially seem that the time-independent $\eta$  could surpass $\eta_m$ for specific temperature scenarios. However, when we consider the effective working conditions governed by the equation:
\begin{equation}
\frac{E_1}{T_1} + \frac{E_3}{T_3} < \frac{E_2}{T_2},
\end{equation}
and adhere to the self-contained condition $E_2 = E_1 + E_3$, we find
\begin{equation}
\eta = \frac{E_1}{E_3} \le \frac{(1/T_2 - 1/T_3)}{(1/T_1 - 1/T_2)} = \frac{(T_3 - T_2)T_1}{T_3(T_2 - T_1)} = \eta_m.
\end{equation}
This demonstrates that the actual COP will never surpass the limitations imposed by the laws of thermodynamics.


\begin{figure}[htbp]
\centering
\includegraphics[width=0.95\linewidth]{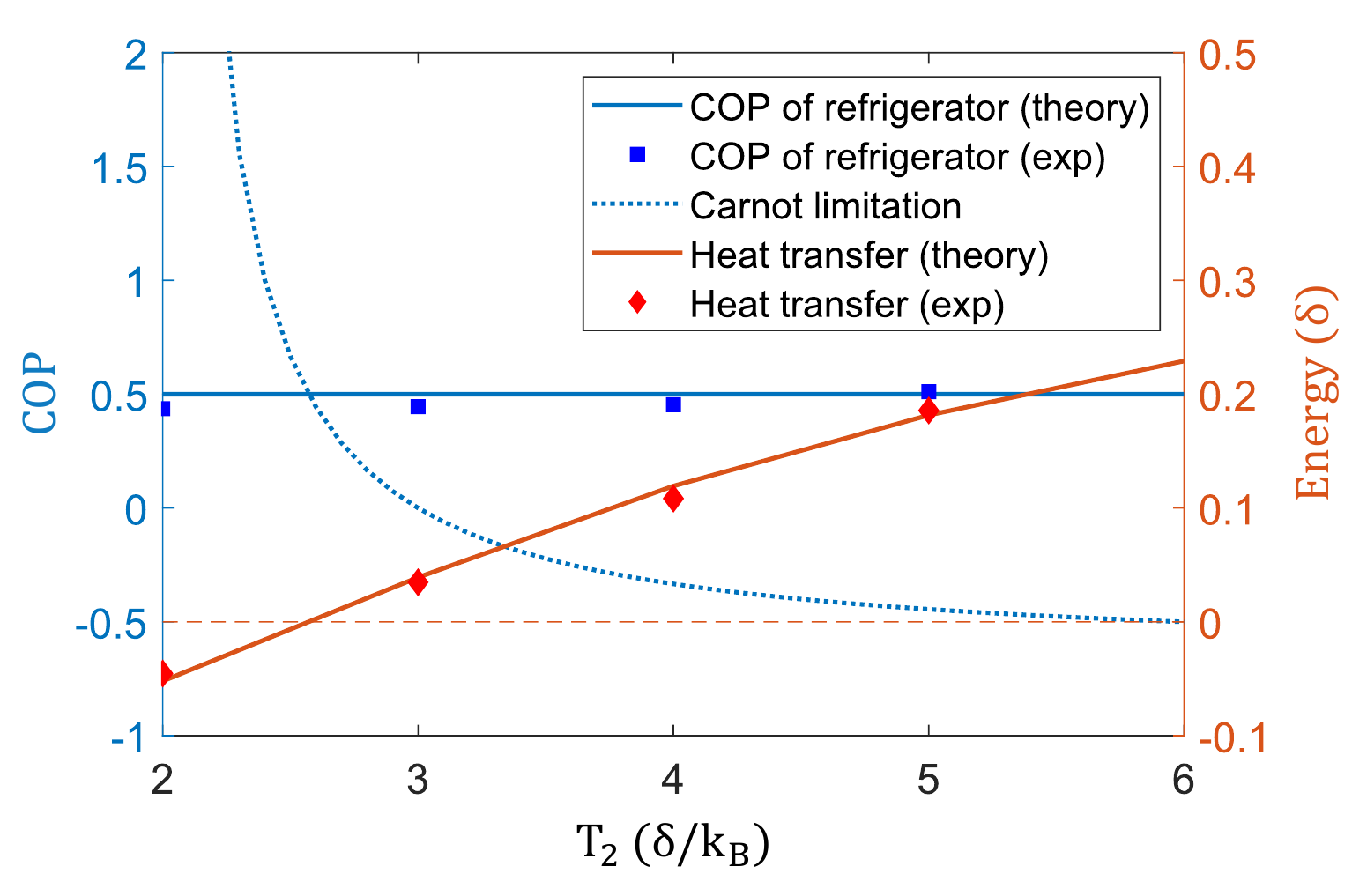}
\caption{COP and heat transfer of the first qubit with different temperatures of the second qubit $T_2$. The blue lines represent the COP with their axis on the left, while the red lines represent the energy with their axis on the right.}
\label{effwork}
\end{figure}

To demonstrate this experimentally, we conducted measurements of the heat transfer of the first qubit and the COP of the refrigerator at various temperatures, as illustrated in Figure~\ref{effwork}. As the temperature increases and the Carnot limitation falls below the constant COP of the refrigerator, the heat transfer diminishes to values below zero. This phenomenon signifies the breakdown or failure of the refrigeration cooling process.

\section{Appendix H: Refrigeration cycle in NMR system}\label{RC}
\begin{figure*}[htbp]
    \centering
    \includegraphics[width=1\linewidth]{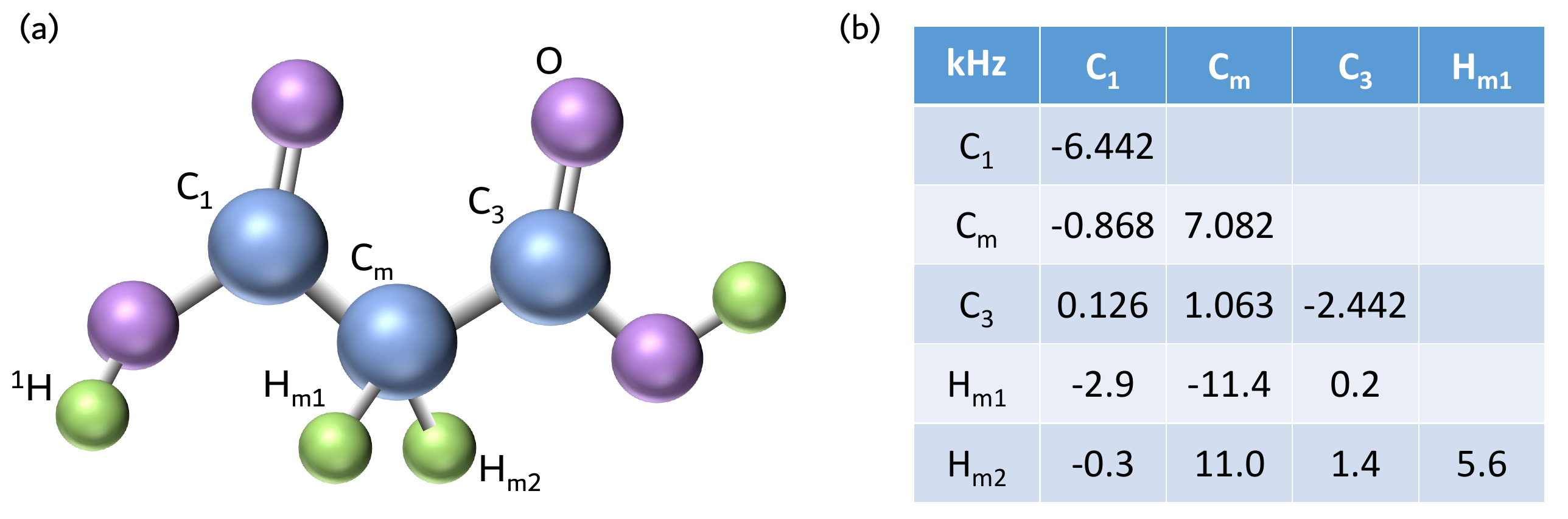}
    \caption{{(a) Structure of the malonic acid molecule. The labeled H$_m$ represents specific hydrogen nuclei of the methylene $^{1}$H pair, while all other hydrogen nuclei are denoted as $^1$H. The refrigeration process is facilitated by the three $^{13}$C nuclei spins, with the abundant $^{1}$H nuclei serving as the thermal bath. (b) Parameters of inter nuclear coupling strength in malonic acid molecule.}}\label{Rf1}
\end{figure*}

The optimal approach for realizing the cooling cycle in NMR involves solid-state systems, where numerous nuclear spins can serve as a high-heat-capacity spin bath. These spins can be selectively coupled or decoupled from an embedded ensemble of spin-labeled isotopomers, the qubits. As an example, a promising candidate for implementing self-contained refrigeration through multiple cycles is the single-crystal malonic acid, illustrated in Figure~\ref{Rf1}. In each cooling cycle, refrigeration is achieved by the three $^{13}$C nuclei spins, C$_1$, C$_m$, and C$_3$, with the abundant $^{1}$H nuclei acting as the thermal bath. Notably, the methylene carbon C$_m$ exhibits a dipolar coupling with the H$_m$ proton of the methylene $^{1}$H pair. The coupling strength is approximately 19 kHz, enabling the H$_m$ proton to function as the ``medium" for establishing effective thermal contact within a duration of 0.1 ms.

\begin{figure*}[htbp]
    \centering
    \includegraphics[width=\linewidth]{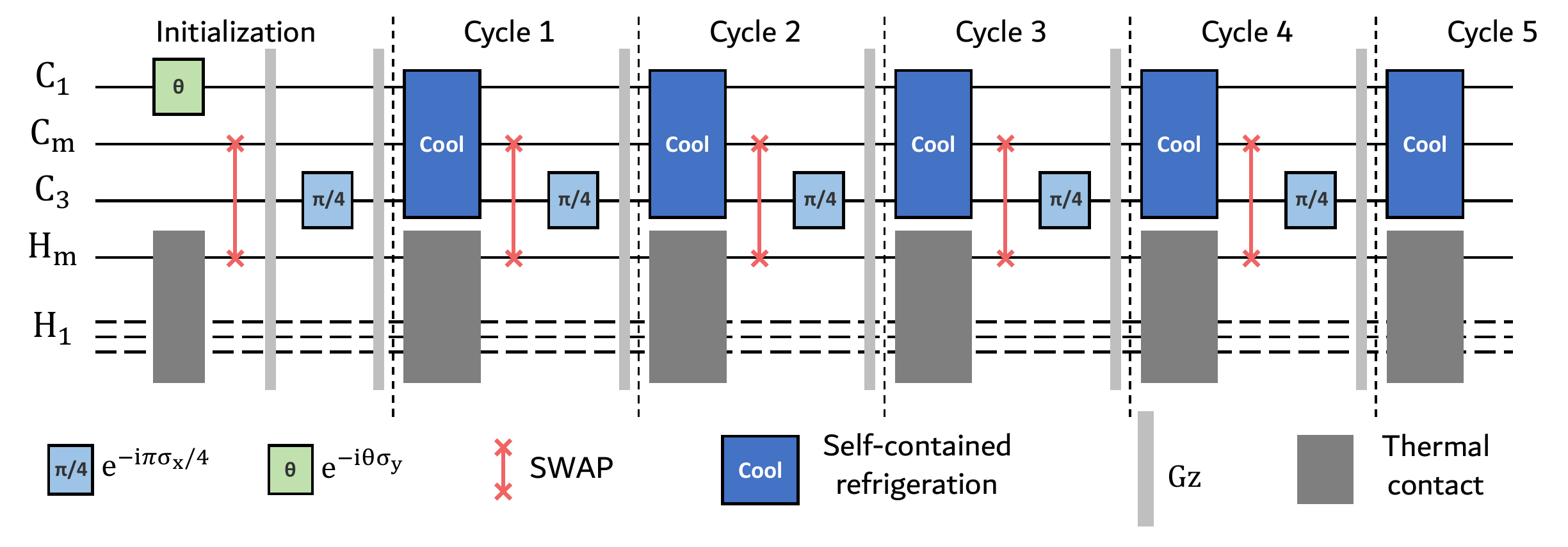}
    \caption{{Quantum circuit for the cooling cycle, consisting of the initialization stage and five self-contained refrigeration cycles. The physical system under consideration is the single-crystal malonic acid in the solid-state NMR setup.}}\label{Rf2}
\end{figure*}

The operation of the cooling cycle is illustrated in Figure~\ref{Rf2}, utilizing three $^{13}$C nuclei spins for refrigeration, with $^{1}$H nuclei serving as the thermal bath for resetting. Without loss of generality, we display five cooling cycles in the figure. The cycle commences with initialization, where the target spin C$_1$ is prepared to the designated temperature through a single-spin rotation ($e^{-i\theta\sigma_y}$) with a tunable parameter $\theta$ for population redistribution, followed by a $z$-gradient field to eliminate unwanted coherence. The second spin C$_m$ is initialized to the temperature of the abundant $^{1}$H thermal bath via thermal contact with $^{1}$H and H$_m$, followed by a SWAP operation between C$_m$ and H$_m$. As the polarization of the $^{1}$H spin is approximately four times higher than that of $^{13}$C spins, the effective temperature of C$_m$ is significantly reduced. The third spin C$_3$ is initialized to effective infinite temperature through a $\pi/4$ rotation and a $z$-gradient field. After this initialization, the temperatures of all three $^{13}$C nuclei spins satisfy the self-contained refrigeration requirement [Eq. (2) in the main text]. It is essential to note that the parameters in this initialization stage are not unique and can be adjusted based on requirements, as long as the final temperature condition is met.

Subsequently, the cooling cycle involves a self-contained refrigeration process and a reset mechanism for C$_m$ and C$_3$. The self-contained refrigeration replicates the process demonstrated in the main text, and we omit its implementation details here. For the reset mechanism of C$_m$ and C$_3$, it resembles the technique used in the initialization stage. C$_m$ is refreshed by swapping with H$_m$, in direct contact with the thermal bath. C$_3$ is refreshed via a $\pi/4$ rotation and a $z$-gradient field, simulating contact with an infinite-temperature thermal bath. Importantly, we aim to minimize changes to the $^{13}$C nuclei spins during the thermal contact between $^{1}$H and H$_m$. This is achieved by rotating the $^{1}$H polarization into the transverse plane and then ``spin-locking" it with a strong, phase-matched radiofrequency field. This technique preserves bulk $^{1}$H polarization and decouples $^{1}$H-$^{13}$C dipolar interactions.

To maximize the number of cooling cycles, it is crucial that the reset timescale of C$_m$ and C$_3$ is significantly shorter than the decoherence time of C$_1$ to preserve its state (temperature) during the reset process. Given the parameters in Figure 1(b), where the coupling between adjacent $^{13}$C nuclei within one molecule is on the order of 1 kHz, the self-contained refrigeration operation can be performed rapidly (approximately 1.5 ms). {The refresh times for C$_m$ and C$_3$ are approximately $40\ \mu$s for the SWAP via cross-polarization and $2.5$ ms using pulse technology, respectively.} Consequently, the duration of one cooling cycle is less than 5 ms, allowing for multiple cycles to occur before the decoherence of the target spin C$_1$ ({longer than 100 ms}).

\begin{figure*}[htbp]
    \centering
    \includegraphics[width=\linewidth]{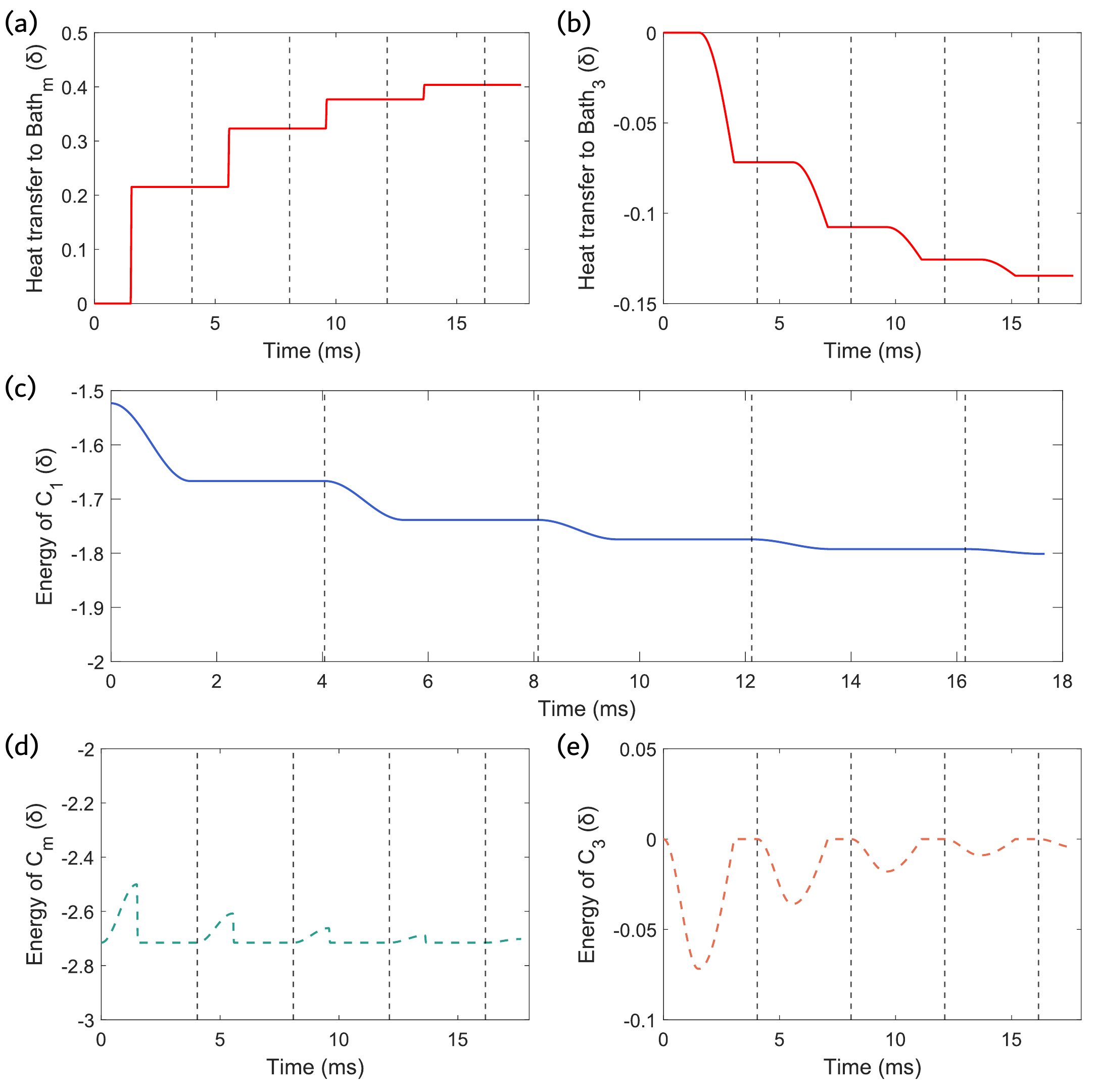}
    \caption{{Heat transfer and changes in internal energy during the five cooling cycles. Following initialization, (a) and (b) illustrate heat transfer to the thermal baths associated with C$_m$ and C$_3$, while (c) to (e) showcase the internal energy changes of C$_1$, C$_m$, and C$_3$, respectively.}}\label{Rf3}
\end{figure*}

We have performed numerical simulations to illustrate the cooling mechanism, with a focus on heat transfer to the thermal baths and the internal energy dynamics of each $^{13}$C nucleus spin during the cycle. Figure~\ref{Rf3} showcases these results, demonstrating the repetitive refreshing of internal energy for C$_m$ [Figure~\ref{Rf3}(d)] and C$_3$ [Figure~\ref{Rf3}(e)], alongside the continuous decrease in internal energy for the target spin C$_1$ [Figure~\ref{Rf3}(c)]. The heat transfer analysis reveals a consistent flow of heat from the higher-temperature thermal bath to the lower-temperature thermal bath, aligning with the desired refrigeration effect.

\section{Appendix I: Populations of two exchanged levels}

\begin{figure}[htbp]
\centering
\includegraphics[width=0.9\linewidth]{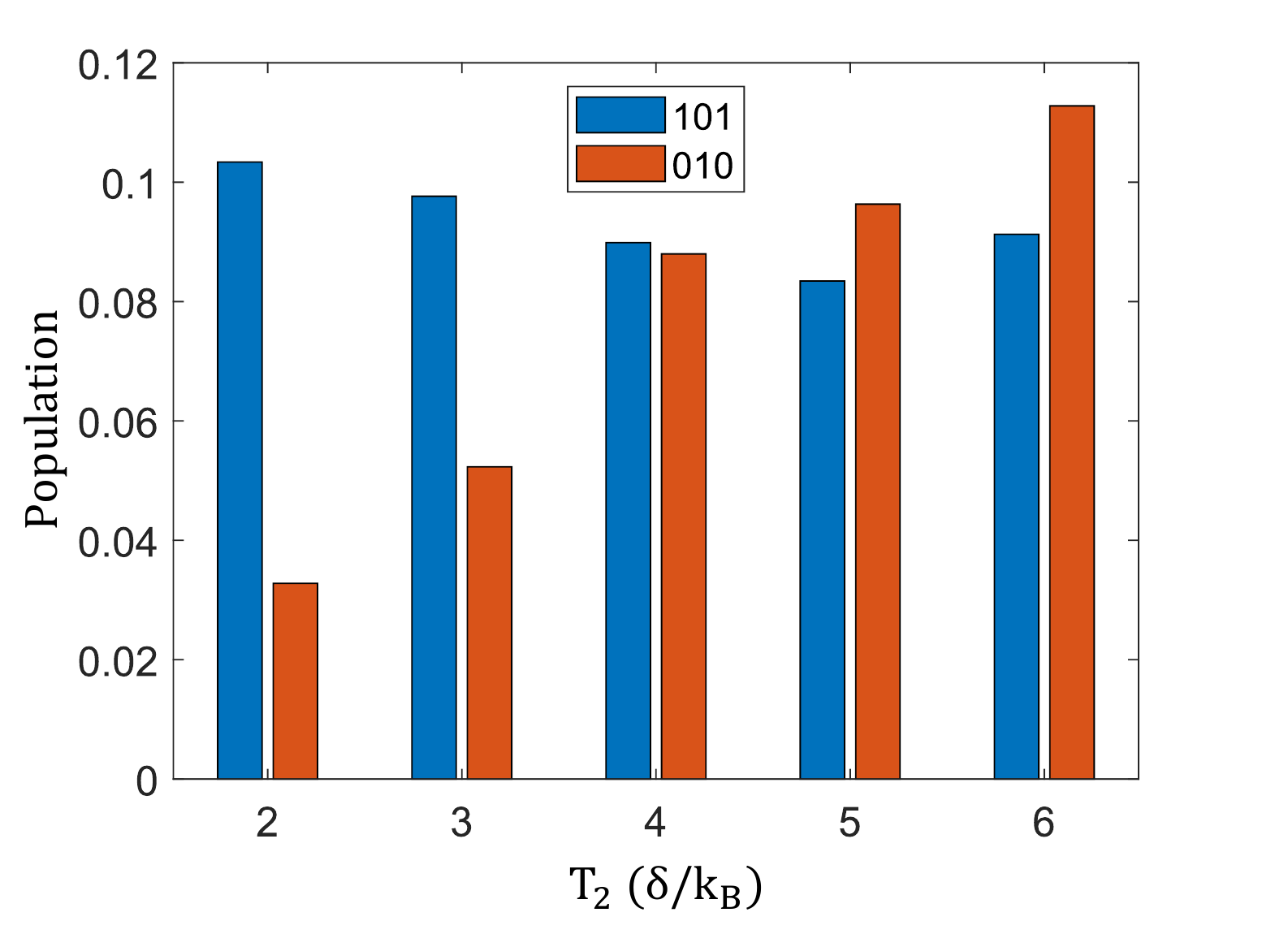}
\caption{Populations of $\vert 010\rangle$ and $\vert 101\rangle$  before exchange at different $T_2$ with   $T_1=2\delta/k_B$ and $T_3=10\delta/k_B$ respectively.}
\label{population}
\end{figure}

The populations of the states $\left|010\right>$ and $\left|101\right>$ play a crucial role in the operation of this self-contained refrigeration. We conducted experimental measurements of the initial  populations of the two states  at different $T_2$, as depicted in  Figure~\ref{population}.
At lower temperatures, the population of the state $\left|101\right>$ is higher than that of the state $\left|010\right>$. As the temperature of the second spin increases, the population of the state $\left|010\right>$ decreases, while the population of the state $\left|101\right>$ increases. When the temperature reaches a sufficiently high level, the population of $\left|010\right>$ surpasses that of $\vert 101\rangle$, signifying a positive heat transfer to the first qubit after the population exchange. This, in turn, indicates the failure of the cooling process.

\section{Appendix J: Embedment of the refrigeration in quantum computation}
This refrigeration can achieve a cooling for some spins. Thus, embedding this refrigeration within a quantum computation is helpful for improving the performance of qubits by decreasing their temperature.  We will give the feasibility of this scheme in this section.

\begin{figure*}[htbp]
    \centering
    \includegraphics[width=\linewidth]{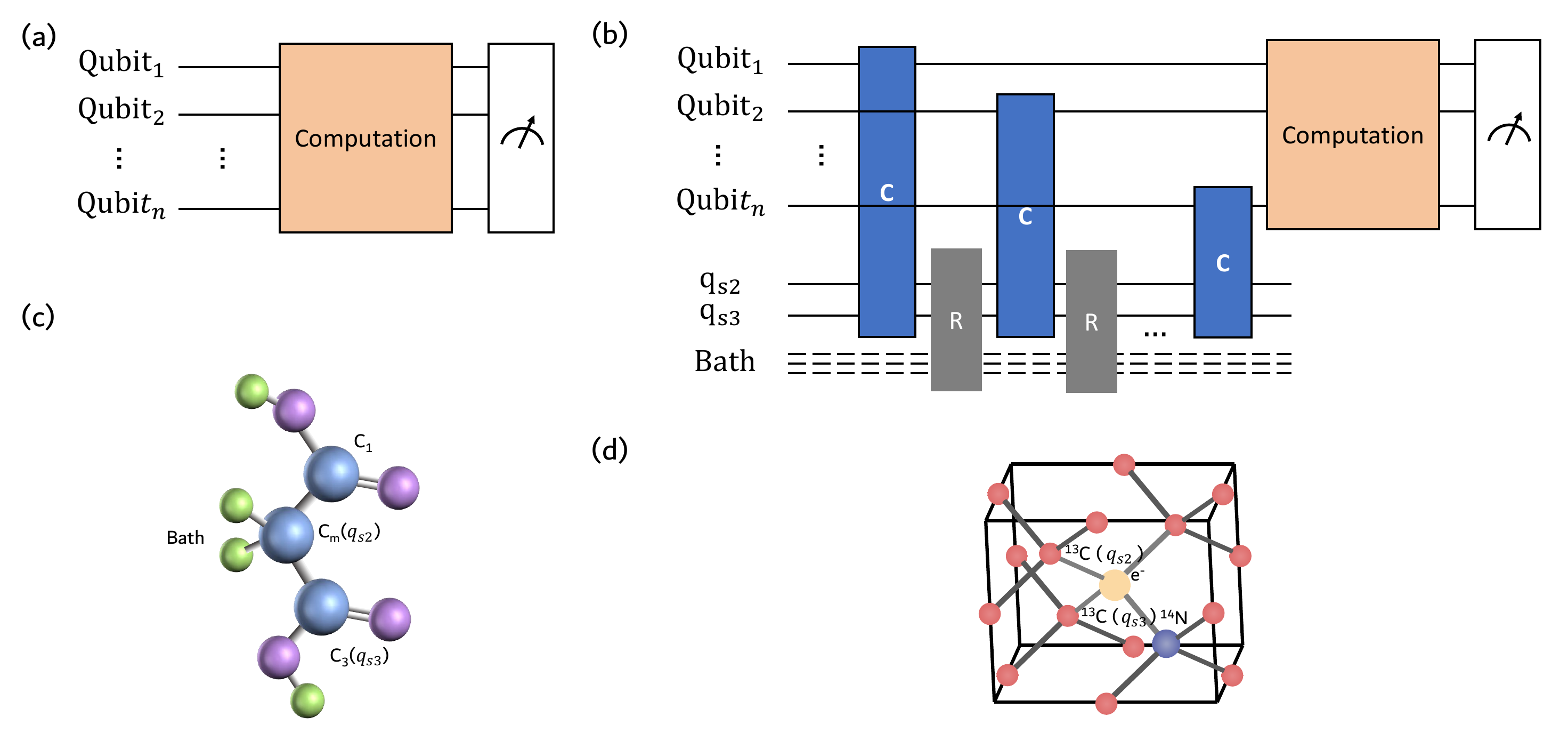}
    \caption{{Schematic representation of integrating the cooling scheme within a larger computational cycle. (a) A typical circuit for quantum computation, involving the initialization, computation, and measurement stages. (b) Integration of qubit cooling before computation, with operations C and R denoting the refrigeration and resetting processes, respectively. (c) Structure of malonic acid molecule, where C$_m$ and C$_3$ can refrigerate C$_1$ for computation and H nuclei labeled by green balls consist the thermal bath. (d) Structure of the NV center in diamond, where the electron spin, $^{14}$N, and $^{13}$C nuclear spins can be utilized for implementing the cooling cycles and subsequent quantum computing tasks.}}\label{Rf6}
\end{figure*}

As depicted in Figure~\ref{Rf6}(a), a typical quantum computation process involves the initial preparation of qubits, followed by the application of operations to facilitate their evolution (computing step), and concludes with measurements to obtain results. Inherent errors may arise during this process, primarily due to qubit impurities and energy-related disturbances, as highlighted in your query.

To counteract these errors, we propose that a modified computational process, incorporating this refrigeration method, as illustrated in Figure~\ref{Rf6}(b). {We apply this cooling cycle to the qubits before they are used for computation.Every cycle involves the application of our refrigeration technique to each qubit, accompanied by subsequent work qubit refreshing. For the refrigeration operation, $q_{s2}$ and $q_{s3}$ contacted with baths are used to cool the target computational qubit. Then they are refreshed by the relaxation in contacting with baths.} This approach effectively diminishes the energy-induced errors, enhancing computational accuracy.

This methodology is particularly applicable to solid-state spin systems. In ensemble systems, for example, molecules and their constituent nuclei serve as a quantum computing framework. Refrigerating these nuclear spins is instrumental in lowering the error rate. The cycle of refrigeration in this system has been discussed in Appendix~\ref{RC}, and the refrigeration to qubit for computation is similar. Consider the case of single-crystal malonic acid, where the three $^{13}$C labeled and $^1$H nuclei can be manipulated to execute this process, as previously depicted in Figure~\ref{Rf1}. By designating the three labeled carbons as C$_1$, C$_m$, and C$_3$, where C$_1$ serves as the computation qubit and C$_m$ and C$_3$ act as $q_{s2}$ and $q_{s3}$ from Figure~\ref{Rf6}(c), and leveraging the $^1$H nuclei as a thermal bath, we can achieve efficient refrigeration.

For single solid spin systems, such as NV centers, the structure comprises one $^{14}$N and multiple (e.g., two) $^{13}$C nuclear spins. The nuclear spins interact with the central electron spin at coupling strengths of 2.16 MHz, 90 kHz, and 414 kHz, respectively [Figure~\ref{Rf6}(d)]. By selecting the $^{14}$N as the computational qubit and utilizing the two $^{13}$C nuclei as $q_{s2}$ and $q_{s3}$ for cooling, we can effectively reduce errors. Moreover, the central electron serves dual purposes: facilitating refrigeration operations and acting as a thermal bath for resetting the $^{13}$C spins. Given the coupling strengths, typical durations for the compression and refresh steps are approximately 0.3 ms and 5 ms, respectively. The nuclear spin's lifetime extends to the order of seconds, permitting multiple iterations of the cooling cycle.

During the cooling cycle, the thermal bath is represented by the electron spin, while the target spin (to be cooled down) is the nuclear spin coupled to the electron spin through hyperfine interactions. In this case, the heat flow dynamics can be achieved using laser cooling. After the refrigeration cycle, the electron spin undergoes re-initialization through optical pumping, effectively removing excess heat and ensuring the thermal bath's readiness for subsequent cooling cycles. Given the short reset time of the electron spin compared to the decoherence time of the nuclear spin, multiple cooling cycles can be implemented for continuous cooling of the target spin.


\begin{thebibliography}{53}%
\makeatletter
\providecommand \@ifxundefined [1]{%
 \@ifx{#1\undefined}
}%
\providecommand \@ifnum [1]{%
 \ifnum #1\expandafter \@firstoftwo
 \else \expandafter \@secondoftwo
 \fi
}%
\providecommand \@ifx [1]{%
 \ifx #1\expandafter \@firstoftwo
 \else \expandafter \@secondoftwo
 \fi
}%
\providecommand \natexlab [1]{#1}%
\providecommand \enquote  [1]{``#1''}%
\providecommand \bibnamefont  [1]{#1}%
\providecommand \bibfnamefont [1]{#1}%
\providecommand \citenamefont [1]{#1}%
\providecommand \href@noop [0]{\@secondoftwo}%
\providecommand \href [0]{\begingroup \@sanitize@url \@href}%
\providecommand \@href[1]{\@@startlink{#1}\@@href}%
\providecommand \@@href[1]{\endgroup#1\@@endlink}%
\providecommand \@sanitize@url [0]{\catcode `\\12\catcode `\$12\catcode
  `\&12\catcode `\#12\catcode `\^12\catcode `\_12\catcode `\%12\relax}%
\providecommand \@@startlink[1]{}%
\providecommand \@@endlink[0]{}%
\providecommand \url  [0]{\begingroup\@sanitize@url \@url }%
\providecommand \@url [1]{\endgroup\@href {#1}{\urlprefix }}%
\providecommand \urlprefix  [0]{URL }%
\providecommand \Eprint [0]{\href }%
\providecommand \doibase [0]{http://dx.doi.org/}%
\providecommand \selectlanguage [0]{\@gobble}%
\providecommand \bibinfo  [0]{\@secondoftwo}%
\providecommand \bibfield  [0]{\@secondoftwo}%
\providecommand \translation [1]{[#1]}%
\providecommand \BibitemOpen [0]{}%
\providecommand \bibitemStop [0]{}%
\providecommand \bibitemNoStop [0]{.\EOS\space}%
\providecommand \EOS [0]{\spacefactor3000\relax}%
\providecommand \BibitemShut  [1]{\csname bibitem#1\endcsname}%
\let\auto@bib@innerbib\@empty
\bibitem [{\citenamefont {Combescot}(2022)}]{combescot2022superconductivity}%
  \BibitemOpen
  \bibfield  {author} {\bibinfo {author} {\bibfnamefont {R.}~\bibnamefont
  {Combescot}},\ }\href {https://books.google.com.hk/books?id=_1tqEAAAQBAJ}
  {\emph {\bibinfo {title} {Superconductivity: An Introduction}}}\ (\bibinfo
  {publisher} {Cambridge University Press},\ \bibinfo {year}
  {2022})\BibitemShut {NoStop}%
\bibitem [{\citenamefont {Davis}\ \emph {et~al.}(1995)\citenamefont {Davis},
  \citenamefont {Mewes}, \citenamefont {Andrews}, \citenamefont {van Druten},
  \citenamefont {Durfee}, \citenamefont {Kurn},\ and\ \citenamefont
  {Ketterle}}]{PhysRevLett.75.3969}%
  \BibitemOpen
  \bibfield  {author} {\bibinfo {author} {\bibfnamefont {K.~B.}\ \bibnamefont
  {Davis}}, \bibinfo {author} {\bibfnamefont {M.~O.}\ \bibnamefont {Mewes}},
  \bibinfo {author} {\bibfnamefont {M.~R.}\ \bibnamefont {Andrews}}, \bibinfo
  {author} {\bibfnamefont {N.~J.}\ \bibnamefont {van Druten}}, \bibinfo
  {author} {\bibfnamefont {D.~S.}\ \bibnamefont {Durfee}}, \bibinfo {author}
  {\bibfnamefont {D.~M.}\ \bibnamefont {Kurn}}, \ and\ \bibinfo {author}
  {\bibfnamefont {W.}~\bibnamefont {Ketterle}},\ }\href {\doibase
  10.1103/PhysRevLett.75.3969} {\bibfield  {journal} {\bibinfo  {journal}
  {Phys. Rev. Lett.}\ }\textbf {\bibinfo {volume} {75}},\ \bibinfo {pages}
  {3969} (\bibinfo {year} {1995})}\BibitemShut {NoStop}%
\bibitem [{\citenamefont {Anderson}\ \emph {et~al.}(1995)\citenamefont
  {Anderson}, \citenamefont {Ensher}, \citenamefont {Matthews}, \citenamefont
  {Wieman},\ and\ \citenamefont {Cornell}}]{doi:10.1126/science.269.5221.198}%
  \BibitemOpen
  \bibfield  {author} {\bibinfo {author} {\bibfnamefont {M.~H.}\ \bibnamefont
  {Anderson}}, \bibinfo {author} {\bibfnamefont {J.~R.}\ \bibnamefont
  {Ensher}}, \bibinfo {author} {\bibfnamefont {M.~R.}\ \bibnamefont
  {Matthews}}, \bibinfo {author} {\bibfnamefont {C.~E.}\ \bibnamefont
  {Wieman}}, \ and\ \bibinfo {author} {\bibfnamefont {E.~A.}\ \bibnamefont
  {Cornell}},\ }\href {\doibase 10.1126/science.269.5221.198} {\bibfield
  {journal} {\bibinfo  {journal} {Science}\ }\textbf {\bibinfo {volume}
  {269}},\ \bibinfo {pages} {198} (\bibinfo {year} {1995})}\BibitemShut
  {NoStop}%
\bibitem [{\citenamefont {Valenzuela}\ \emph {et~al.}(2006)\citenamefont
  {Valenzuela}, \citenamefont {Oliver}, \citenamefont {Berns}, \citenamefont
  {Berggren}, \citenamefont {Levitov},\ and\ \citenamefont
  {Orlando}}]{doi:10.1126/science.1134008}%
  \BibitemOpen
  \bibfield  {author} {\bibinfo {author} {\bibfnamefont {S.~O.}\ \bibnamefont
  {Valenzuela}}, \bibinfo {author} {\bibfnamefont {W.~D.}\ \bibnamefont
  {Oliver}}, \bibinfo {author} {\bibfnamefont {D.~M.}\ \bibnamefont {Berns}},
  \bibinfo {author} {\bibfnamefont {K.~K.}\ \bibnamefont {Berggren}}, \bibinfo
  {author} {\bibfnamefont {L.~S.}\ \bibnamefont {Levitov}}, \ and\ \bibinfo
  {author} {\bibfnamefont {T.~P.}\ \bibnamefont {Orlando}},\ }\href {\doibase
  10.1126/science.1134008} {\bibfield  {journal} {\bibinfo  {journal}
  {Science}\ }\textbf {\bibinfo {volume} {314}},\ \bibinfo {pages} {1589}
  (\bibinfo {year} {2006})}\BibitemShut {NoStop}%
\bibitem [{\citenamefont {Tan}\ \emph {et~al.}(2017)\citenamefont {Tan},
  \citenamefont {Partanen}, \citenamefont {Lake}, \citenamefont {Govenius},
  \citenamefont {Masuda},\ and\ \citenamefont {M{\"o}tt{\"o}nen}}]{Tan2017}%
  \BibitemOpen
  \bibfield  {author} {\bibinfo {author} {\bibfnamefont {K.~Y.}\ \bibnamefont
  {Tan}}, \bibinfo {author} {\bibfnamefont {M.}~\bibnamefont {Partanen}},
  \bibinfo {author} {\bibfnamefont {R.~E.}\ \bibnamefont {Lake}}, \bibinfo
  {author} {\bibfnamefont {J.}~\bibnamefont {Govenius}}, \bibinfo {author}
  {\bibfnamefont {S.}~\bibnamefont {Masuda}}, \ and\ \bibinfo {author}
  {\bibfnamefont {M.}~\bibnamefont {M{\"o}tt{\"o}nen}},\ }\href {\doibase
  10.1038/ncomms15189} {\bibfield  {journal} {\bibinfo  {journal} {Nat.
  Commun.}\ }\textbf {\bibinfo {volume} {8}},\ \bibinfo {pages} {15189}
  (\bibinfo {year} {2017})}\BibitemShut {NoStop}%
\bibitem [{\citenamefont {Nakamura}\ \emph {et~al.}(1999)\citenamefont
  {Nakamura}, \citenamefont {Pashkin},\ and\ \citenamefont
  {Tsai}}]{Nakamura1999}%
  \BibitemOpen
  \bibfield  {author} {\bibinfo {author} {\bibfnamefont {Y.}~\bibnamefont
  {Nakamura}}, \bibinfo {author} {\bibfnamefont {Y.~A.}\ \bibnamefont
  {Pashkin}}, \ and\ \bibinfo {author} {\bibfnamefont {J.~S.}\ \bibnamefont
  {Tsai}},\ }\href {\doibase 10.1038/19718} {\bibfield  {journal} {\bibinfo
  {journal} {Nature}\ }\textbf {\bibinfo {volume} {398}},\ \bibinfo {pages}
  {786} (\bibinfo {year} {1999})}\BibitemShut {NoStop}%
\bibitem [{\citenamefont {Martinis}\ \emph {et~al.}(2002)\citenamefont
  {Martinis}, \citenamefont {Nam}, \citenamefont {Aumentado},\ and\
  \citenamefont {Urbina}}]{PhysRevLett.89.117901}%
  \BibitemOpen
  \bibfield  {author} {\bibinfo {author} {\bibfnamefont {J.~M.}\ \bibnamefont
  {Martinis}}, \bibinfo {author} {\bibfnamefont {S.}~\bibnamefont {Nam}},
  \bibinfo {author} {\bibfnamefont {J.}~\bibnamefont {Aumentado}}, \ and\
  \bibinfo {author} {\bibfnamefont {C.}~\bibnamefont {Urbina}},\ }\href
  {\doibase 10.1103/PhysRevLett.89.117901} {\bibfield  {journal} {\bibinfo
  {journal} {Phys. Rev. Lett.}\ }\textbf {\bibinfo {volume} {89}},\ \bibinfo
  {pages} {117901} (\bibinfo {year} {2002})}\BibitemShut {NoStop}%
\bibitem [{\citenamefont {Bal}\ \emph {et~al.}(2012)\citenamefont {Bal},
  \citenamefont {Deng}, \citenamefont {Orgiazzi}, \citenamefont {Ong},\ and\
  \citenamefont {Lupascu}}]{Bal2012}%
  \BibitemOpen
  \bibfield  {author} {\bibinfo {author} {\bibfnamefont {M.}~\bibnamefont
  {Bal}}, \bibinfo {author} {\bibfnamefont {C.}~\bibnamefont {Deng}}, \bibinfo
  {author} {\bibfnamefont {J.-L.}\ \bibnamefont {Orgiazzi}}, \bibinfo {author}
  {\bibfnamefont {F.~R.}\ \bibnamefont {Ong}}, \ and\ \bibinfo {author}
  {\bibfnamefont {A.}~\bibnamefont {Lupascu}},\ }\href {\doibase
  10.1038/ncomms2332} {\bibfield  {journal} {\bibinfo  {journal} {Nat.
  Commun.}\ }\textbf {\bibinfo {volume} {3}},\ \bibinfo {pages} {1324}
  (\bibinfo {year} {2012})}\BibitemShut {NoStop}%
\bibitem [{\citenamefont {Degen}\ \emph {et~al.}(2017)\citenamefont {Degen},
  \citenamefont {Reinhard},\ and\ \citenamefont
  {Cappellaro}}]{RevModPhys.89.035002}%
  \BibitemOpen
  \bibfield  {author} {\bibinfo {author} {\bibfnamefont {C.~L.}\ \bibnamefont
  {Degen}}, \bibinfo {author} {\bibfnamefont {F.}~\bibnamefont {Reinhard}}, \
  and\ \bibinfo {author} {\bibfnamefont {P.}~\bibnamefont {Cappellaro}},\
  }\href {\doibase 10.1103/RevModPhys.89.035002} {\bibfield  {journal}
  {\bibinfo  {journal} {Rev. Mod. Phys.}\ }\textbf {\bibinfo {volume} {89}},\
  \bibinfo {pages} {035002} (\bibinfo {year} {2017})}\BibitemShut {NoStop}%
\bibitem [{\citenamefont {Cheng}\ \emph {et~al.}(2023)\citenamefont {Cheng},
  \citenamefont {Deng}, \citenamefont {Gu}, \citenamefont {He}, \citenamefont
  {Hu}, \citenamefont {Huang}, \citenamefont {Li}, \citenamefont {Lin},
  \citenamefont {Lu}, \citenamefont {Lu} \emph {et~al.}}]{cheng2023noisy}%
  \BibitemOpen
  \bibfield  {author} {\bibinfo {author} {\bibfnamefont {B.}~\bibnamefont
  {Cheng}}, \bibinfo {author} {\bibfnamefont {X.-H.}\ \bibnamefont {Deng}},
  \bibinfo {author} {\bibfnamefont {X.}~\bibnamefont {Gu}}, \bibinfo {author}
  {\bibfnamefont {Y.}~\bibnamefont {He}}, \bibinfo {author} {\bibfnamefont
  {G.}~\bibnamefont {Hu}}, \bibinfo {author} {\bibfnamefont {P.}~\bibnamefont
  {Huang}}, \bibinfo {author} {\bibfnamefont {J.}~\bibnamefont {Li}}, \bibinfo
  {author} {\bibfnamefont {B.-C.}\ \bibnamefont {Lin}}, \bibinfo {author}
  {\bibfnamefont {D.}~\bibnamefont {Lu}}, \bibinfo {author} {\bibfnamefont
  {Y.}~\bibnamefont {Lu}},  \emph {et~al.},\ }\href
  {https://doi.org/10.1007/s11467-022-1249-z} {\bibfield  {journal} {\bibinfo
  {journal} {Front. Phys.}\ }\textbf {\bibinfo {volume} {18}},\ \bibinfo
  {pages} {21308} (\bibinfo {year} {2023})}\BibitemShut {NoStop}%
\bibitem [{\citenamefont {Chitambar}\ and\ \citenamefont
  {Gour}(2019)}]{RevModPhys.91.025001}%
  \BibitemOpen
  \bibfield  {author} {\bibinfo {author} {\bibfnamefont {E.}~\bibnamefont
  {Chitambar}}\ and\ \bibinfo {author} {\bibfnamefont {G.}~\bibnamefont
  {Gour}},\ }\href {\doibase 10.1103/RevModPhys.91.025001} {\bibfield
  {journal} {\bibinfo  {journal} {Rev. Mod. Phys.}\ }\textbf {\bibinfo {volume}
  {91}},\ \bibinfo {pages} {025001} (\bibinfo {year} {2019})}\BibitemShut
  {NoStop}%
\bibitem [{\citenamefont {Parrondo}\ \emph {et~al.}(2015)\citenamefont
  {Parrondo}, \citenamefont {Horowitz},\ and\ \citenamefont
  {Sagawa}}]{Parrondo2015}%
  \BibitemOpen
  \bibfield  {author} {\bibinfo {author} {\bibfnamefont {J.~M.~R.}\
  \bibnamefont {Parrondo}}, \bibinfo {author} {\bibfnamefont {J.~M.}\
  \bibnamefont {Horowitz}}, \ and\ \bibinfo {author} {\bibfnamefont
  {T.}~\bibnamefont {Sagawa}},\ }\href {\doibase 10.1038/nphys3230} {\bibfield
  {journal} {\bibinfo  {journal} {Nat. Phys.}\ }\textbf {\bibinfo {volume}
  {11}},\ \bibinfo {pages} {131} (\bibinfo {year} {2015})}\BibitemShut
  {NoStop}%
\bibitem [{\citenamefont {Goold}\ \emph {et~al.}(2016)\citenamefont {Goold},
  \citenamefont {Huber}, \citenamefont {Riera}, \citenamefont {del Rio},\ and\
  \citenamefont {Skrzypczyk}}]{Goold_2016}%
  \BibitemOpen
  \bibfield  {author} {\bibinfo {author} {\bibfnamefont {J.}~\bibnamefont
  {Goold}}, \bibinfo {author} {\bibfnamefont {M.}~\bibnamefont {Huber}},
  \bibinfo {author} {\bibfnamefont {A.}~\bibnamefont {Riera}}, \bibinfo
  {author} {\bibfnamefont {L.}~\bibnamefont {del Rio}}, \ and\ \bibinfo
  {author} {\bibfnamefont {P.}~\bibnamefont {Skrzypczyk}},\ }\href {\doibase
  10.1088/1751-8113/49/14/143001} {\bibfield  {journal} {\bibinfo  {journal}
  {J. Phys. A Math. Theor.}\ }\textbf {\bibinfo {volume} {49}},\ \bibinfo
  {pages} {143001} (\bibinfo {year} {2016})}\BibitemShut {NoStop}%
\bibitem [{\citenamefont {Popescu}(2014)}]{Popescu2014}%
  \BibitemOpen
  \bibfield  {author} {\bibinfo {author} {\bibfnamefont {S.}~\bibnamefont
  {Popescu}},\ }\href {\doibase 10.1038/nphys2916} {\bibfield  {journal}
  {\bibinfo  {journal} {Nat. Phys.}\ }\textbf {\bibinfo {volume} {10}},\
  \bibinfo {pages} {264} (\bibinfo {year} {2014})}\BibitemShut {NoStop}%
\bibitem [{\citenamefont {Liu}\ \emph {et~al.}(2022)\citenamefont {Liu},
  \citenamefont {Ebler},\ and\ \citenamefont
  {Dahlsten}}]{PhysRevLett.129.230604}%
  \BibitemOpen
  \bibfield  {author} {\bibinfo {author} {\bibfnamefont {X.}~\bibnamefont
  {Liu}}, \bibinfo {author} {\bibfnamefont {D.}~\bibnamefont {Ebler}}, \ and\
  \bibinfo {author} {\bibfnamefont {O.}~\bibnamefont {Dahlsten}},\ }\href
  {\doibase 10.1103/PhysRevLett.129.230604} {\bibfield  {journal} {\bibinfo
  {journal} {Phys. Rev. Lett.}\ }\textbf {\bibinfo {volume} {129}},\ \bibinfo
  {pages} {230604} (\bibinfo {year} {2022})}\BibitemShut {NoStop}%
\bibitem [{\citenamefont {Skrzypczyk}\ \emph {et~al.}(2011)\citenamefont
  {Skrzypczyk}, \citenamefont {Brunner}, \citenamefont {Linden},\ and\
  \citenamefont {Popescu}}]{Skrzypczyk_2011}%
  \BibitemOpen
  \bibfield  {author} {\bibinfo {author} {\bibfnamefont {P.}~\bibnamefont
  {Skrzypczyk}}, \bibinfo {author} {\bibfnamefont {N.}~\bibnamefont {Brunner}},
  \bibinfo {author} {\bibfnamefont {N.}~\bibnamefont {Linden}}, \ and\ \bibinfo
  {author} {\bibfnamefont {S.}~\bibnamefont {Popescu}},\ }\href {\doibase
  10.1088/1751-8113/44/49/492002} {\bibfield  {journal} {\bibinfo  {journal}
  {J. Phys. A Math. Theor.}\ }\textbf {\bibinfo {volume} {44}},\ \bibinfo
  {pages} {492002} (\bibinfo {year} {2011})}\BibitemShut {NoStop}%
\bibitem [{\citenamefont {Palao}\ \emph {et~al.}(2001)\citenamefont {Palao},
  \citenamefont {Kosloff},\ and\ \citenamefont {Gordon}}]{PhysRevE.64.056130}%
  \BibitemOpen
  \bibfield  {author} {\bibinfo {author} {\bibfnamefont {J.~P.}\ \bibnamefont
  {Palao}}, \bibinfo {author} {\bibfnamefont {R.}~\bibnamefont {Kosloff}}, \
  and\ \bibinfo {author} {\bibfnamefont {J.~M.}\ \bibnamefont {Gordon}},\
  }\href {\doibase 10.1103/PhysRevE.64.056130} {\bibfield  {journal} {\bibinfo
  {journal} {Phys. Rev. E}\ }\textbf {\bibinfo {volume} {64}},\ \bibinfo
  {pages} {056130} (\bibinfo {year} {2001})}\BibitemShut {NoStop}%
\bibitem [{\citenamefont {Felce}\ and\ \citenamefont
  {Vedral}(2020)}]{PhysRevLett.125.070603}%
  \BibitemOpen
  \bibfield  {author} {\bibinfo {author} {\bibfnamefont {D.}~\bibnamefont
  {Felce}}\ and\ \bibinfo {author} {\bibfnamefont {V.}~\bibnamefont {Vedral}},\
  }\href {\doibase 10.1103/PhysRevLett.125.070603} {\bibfield  {journal}
  {\bibinfo  {journal} {Phys. Rev. Lett.}\ }\textbf {\bibinfo {volume} {125}},\
  \bibinfo {pages} {070603} (\bibinfo {year} {2020})}\BibitemShut {NoStop}%
\bibitem [{\citenamefont {Nie}\ \emph {et~al.}(2022)\citenamefont {Nie},
  \citenamefont {Zhu}, \citenamefont {Huang}, \citenamefont {Tang},
  \citenamefont {Long}, \citenamefont {Lin}, \citenamefont {Tian},
  \citenamefont {Qiu}, \citenamefont {Xi}, \citenamefont {Yang}, \citenamefont
  {Li}, \citenamefont {Dong}, \citenamefont {Xin},\ and\ \citenamefont
  {Lu}}]{PhysRevLett.129.100603}%
  \BibitemOpen
  \bibfield  {author} {\bibinfo {author} {\bibfnamefont {X.}~\bibnamefont
  {Nie}}, \bibinfo {author} {\bibfnamefont {X.}~\bibnamefont {Zhu}}, \bibinfo
  {author} {\bibfnamefont {K.}~\bibnamefont {Huang}}, \bibinfo {author}
  {\bibfnamefont {K.}~\bibnamefont {Tang}}, \bibinfo {author} {\bibfnamefont
  {X.}~\bibnamefont {Long}}, \bibinfo {author} {\bibfnamefont {Z.}~\bibnamefont
  {Lin}}, \bibinfo {author} {\bibfnamefont {Y.}~\bibnamefont {Tian}}, \bibinfo
  {author} {\bibfnamefont {C.}~\bibnamefont {Qiu}}, \bibinfo {author}
  {\bibfnamefont {C.}~\bibnamefont {Xi}}, \bibinfo {author} {\bibfnamefont
  {X.}~\bibnamefont {Yang}}, \bibinfo {author} {\bibfnamefont {J.}~\bibnamefont
  {Li}}, \bibinfo {author} {\bibfnamefont {Y.}~\bibnamefont {Dong}}, \bibinfo
  {author} {\bibfnamefont {T.}~\bibnamefont {Xin}}, \ and\ \bibinfo {author}
  {\bibfnamefont {D.}~\bibnamefont {Lu}},\ }\href {\doibase
  10.1103/PhysRevLett.129.100603} {\bibfield  {journal} {\bibinfo  {journal}
  {Phys. Rev. Lett.}\ }\textbf {\bibinfo {volume} {129}},\ \bibinfo {pages}
  {100603} (\bibinfo {year} {2022})}\BibitemShut {NoStop}%
\bibitem [{\citenamefont {Buffoni}\ \emph {et~al.}(2019)\citenamefont
  {Buffoni}, \citenamefont {Solfanelli}, \citenamefont {Verrucchi},
  \citenamefont {Cuccoli},\ and\ \citenamefont
  {Campisi}}]{PhysRevLett.122.070603}%
  \BibitemOpen
  \bibfield  {author} {\bibinfo {author} {\bibfnamefont {L.}~\bibnamefont
  {Buffoni}}, \bibinfo {author} {\bibfnamefont {A.}~\bibnamefont {Solfanelli}},
  \bibinfo {author} {\bibfnamefont {P.}~\bibnamefont {Verrucchi}}, \bibinfo
  {author} {\bibfnamefont {A.}~\bibnamefont {Cuccoli}}, \ and\ \bibinfo
  {author} {\bibfnamefont {M.}~\bibnamefont {Campisi}},\ }\href {\doibase
  10.1103/PhysRevLett.122.070603} {\bibfield  {journal} {\bibinfo  {journal}
  {Phys. Rev. Lett.}\ }\textbf {\bibinfo {volume} {122}},\ \bibinfo {pages}
  {070603} (\bibinfo {year} {2019})}\BibitemShut {NoStop}%
\bibitem [{\citenamefont {Correa}\ \emph {et~al.}(2014)\citenamefont {Correa},
  \citenamefont {Palao}, \citenamefont {Alonso},\ and\ \citenamefont
  {Adesso}}]{Correa2014}%
  \BibitemOpen
  \bibfield  {author} {\bibinfo {author} {\bibfnamefont {L.~A.}\ \bibnamefont
  {Correa}}, \bibinfo {author} {\bibfnamefont {J.~P.}\ \bibnamefont {Palao}},
  \bibinfo {author} {\bibfnamefont {D.}~\bibnamefont {Alonso}}, \ and\ \bibinfo
  {author} {\bibfnamefont {G.}~\bibnamefont {Adesso}},\ }\href {\doibase
  10.1038/srep03949} {\bibfield  {journal} {\bibinfo  {journal} {Sci. Rep.}\
  }\textbf {\bibinfo {volume} {4}},\ \bibinfo {pages} {3949} (\bibinfo {year}
  {2014})}\BibitemShut {NoStop}%
\bibitem [{\citenamefont {Steeneken}\ \emph {et~al.}(2011)\citenamefont
  {Steeneken}, \citenamefont {Le~Phan}, \citenamefont {Goossens}, \citenamefont
  {Koops}, \citenamefont {Brom}, \citenamefont {van~der Avoort},\ and\
  \citenamefont {van Beek}}]{Steeneken2011}%
  \BibitemOpen
  \bibfield  {author} {\bibinfo {author} {\bibfnamefont {P.~G.}\ \bibnamefont
  {Steeneken}}, \bibinfo {author} {\bibfnamefont {K.}~\bibnamefont {Le~Phan}},
  \bibinfo {author} {\bibfnamefont {M.~J.}\ \bibnamefont {Goossens}}, \bibinfo
  {author} {\bibfnamefont {G.~E.~J.}\ \bibnamefont {Koops}}, \bibinfo {author}
  {\bibfnamefont {G.~J. A.~M.}\ \bibnamefont {Brom}}, \bibinfo {author}
  {\bibfnamefont {C.}~\bibnamefont {van~der Avoort}}, \ and\ \bibinfo {author}
  {\bibfnamefont {J.~T.~M.}\ \bibnamefont {van Beek}},\ }\href {\doibase
  10.1038/nphys1871} {\bibfield  {journal} {\bibinfo  {journal} {Nat. Phys.}\
  }\textbf {\bibinfo {volume} {7}},\ \bibinfo {pages} {354} (\bibinfo {year}
  {2011})}\BibitemShut {NoStop}%
\bibitem [{\citenamefont {Mari}\ and\ \citenamefont
  {Eisert}(2012)}]{PhysRevLett.108.120602}%
  \BibitemOpen
  \bibfield  {author} {\bibinfo {author} {\bibfnamefont {A.}~\bibnamefont
  {Mari}}\ and\ \bibinfo {author} {\bibfnamefont {J.}~\bibnamefont {Eisert}},\
  }\href {\doibase 10.1103/PhysRevLett.108.120602} {\bibfield  {journal}
  {\bibinfo  {journal} {Phys. Rev. Lett.}\ }\textbf {\bibinfo {volume} {108}},\
  \bibinfo {pages} {120602} (\bibinfo {year} {2012})}\BibitemShut {NoStop}%
\bibitem [{\citenamefont {Maslennikov}\ \emph {et~al.}(2019)\citenamefont
  {Maslennikov}, \citenamefont {Ding}, \citenamefont {Habl{\"u}tzel},
  \citenamefont {Gan}, \citenamefont {Roulet}, \citenamefont {Nimmrichter},
  \citenamefont {Dai}, \citenamefont {Scarani},\ and\ \citenamefont
  {Matsukevich}}]{Maslennikov2019}%
  \BibitemOpen
  \bibfield  {author} {\bibinfo {author} {\bibfnamefont {G.}~\bibnamefont
  {Maslennikov}}, \bibinfo {author} {\bibfnamefont {S.}~\bibnamefont {Ding}},
  \bibinfo {author} {\bibfnamefont {R.}~\bibnamefont {Habl{\"u}tzel}}, \bibinfo
  {author} {\bibfnamefont {J.}~\bibnamefont {Gan}}, \bibinfo {author}
  {\bibfnamefont {A.}~\bibnamefont {Roulet}}, \bibinfo {author} {\bibfnamefont
  {S.}~\bibnamefont {Nimmrichter}}, \bibinfo {author} {\bibfnamefont
  {J.}~\bibnamefont {Dai}}, \bibinfo {author} {\bibfnamefont {V.}~\bibnamefont
  {Scarani}}, \ and\ \bibinfo {author} {\bibfnamefont {D.}~\bibnamefont
  {Matsukevich}},\ }\href {\doibase 10.1038/s41467-018-08090-0} {\bibfield
  {journal} {\bibinfo  {journal} {Nat. Commun.}\ }\textbf {\bibinfo {volume}
  {10}},\ \bibinfo {pages} {202} (\bibinfo {year} {2019})}\BibitemShut
  {NoStop}%
\bibitem [{\citenamefont {Peterson}\ \emph {et~al.}(2019)\citenamefont
  {Peterson}, \citenamefont {Batalh\~ao}, \citenamefont {Herrera},
  \citenamefont {Souza}, \citenamefont {Sarthour}, \citenamefont {Oliveira},\
  and\ \citenamefont {Serra}}]{PhysRevLett.123.240601}%
  \BibitemOpen
  \bibfield  {author} {\bibinfo {author} {\bibfnamefont {J.~P.~S.}\
  \bibnamefont {Peterson}}, \bibinfo {author} {\bibfnamefont {T.~B.}\
  \bibnamefont {Batalh\~ao}}, \bibinfo {author} {\bibfnamefont
  {M.}~\bibnamefont {Herrera}}, \bibinfo {author} {\bibfnamefont {A.~M.}\
  \bibnamefont {Souza}}, \bibinfo {author} {\bibfnamefont {R.~S.}\ \bibnamefont
  {Sarthour}}, \bibinfo {author} {\bibfnamefont {I.~S.}\ \bibnamefont
  {Oliveira}}, \ and\ \bibinfo {author} {\bibfnamefont {R.~M.}\ \bibnamefont
  {Serra}},\ }\href {\doibase 10.1103/PhysRevLett.123.240601} {\bibfield
  {journal} {\bibinfo  {journal} {Phys. Rev. Lett.}\ }\textbf {\bibinfo
  {volume} {123}},\ \bibinfo {pages} {240601} (\bibinfo {year}
  {2019})}\BibitemShut {NoStop}%
\bibitem [{\citenamefont {Camati}\ \emph {et~al.}(2019)\citenamefont {Camati},
  \citenamefont {Santos},\ and\ \citenamefont {Serra}}]{PhysRevA.99.062103}%
  \BibitemOpen
  \bibfield  {author} {\bibinfo {author} {\bibfnamefont {P.~A.}\ \bibnamefont
  {Camati}}, \bibinfo {author} {\bibfnamefont {J.~F.~G.}\ \bibnamefont
  {Santos}}, \ and\ \bibinfo {author} {\bibfnamefont {R.~M.}\ \bibnamefont
  {Serra}},\ }\href {\doibase 10.1103/PhysRevA.99.062103} {\bibfield  {journal}
  {\bibinfo  {journal} {Phys. Rev. A}\ }\textbf {\bibinfo {volume} {99}},\
  \bibinfo {pages} {062103} (\bibinfo {year} {2019})}\BibitemShut {NoStop}%
\bibitem [{\citenamefont {Skrzypczyk}\ \emph {et~al.}(2014)\citenamefont
  {Skrzypczyk}, \citenamefont {Short},\ and\ \citenamefont
  {Popescu}}]{Skrzypczyk2014}%
  \BibitemOpen
  \bibfield  {author} {\bibinfo {author} {\bibfnamefont {P.}~\bibnamefont
  {Skrzypczyk}}, \bibinfo {author} {\bibfnamefont {A.~J.}\ \bibnamefont
  {Short}}, \ and\ \bibinfo {author} {\bibfnamefont {S.}~\bibnamefont
  {Popescu}},\ }\href {\doibase 10.1038/ncomms5185} {\bibfield  {journal}
  {\bibinfo  {journal} {Nat. Commun.}\ }\textbf {\bibinfo {volume} {5}},\
  \bibinfo {pages} {4185} (\bibinfo {year} {2014})}\BibitemShut {NoStop}%
\bibitem [{\citenamefont {Ji}\ \emph {et~al.}(2022)\citenamefont {Ji},
  \citenamefont {Chai}, \citenamefont {Wang}, \citenamefont {Guo},
  \citenamefont {Rong}, \citenamefont {Shi}, \citenamefont {Ren}, \citenamefont
  {Wang},\ and\ \citenamefont {Du}}]{PhysRevLett.128.090602}%
  \BibitemOpen
  \bibfield  {author} {\bibinfo {author} {\bibfnamefont {W.}~\bibnamefont
  {Ji}}, \bibinfo {author} {\bibfnamefont {Z.}~\bibnamefont {Chai}}, \bibinfo
  {author} {\bibfnamefont {M.}~\bibnamefont {Wang}}, \bibinfo {author}
  {\bibfnamefont {Y.}~\bibnamefont {Guo}}, \bibinfo {author} {\bibfnamefont
  {X.}~\bibnamefont {Rong}}, \bibinfo {author} {\bibfnamefont {F.}~\bibnamefont
  {Shi}}, \bibinfo {author} {\bibfnamefont {C.}~\bibnamefont {Ren}}, \bibinfo
  {author} {\bibfnamefont {Y.}~\bibnamefont {Wang}}, \ and\ \bibinfo {author}
  {\bibfnamefont {J.}~\bibnamefont {Du}},\ }\href {\doibase
  10.1103/PhysRevLett.128.090602} {\bibfield  {journal} {\bibinfo  {journal}
  {Phys. Rev. Lett.}\ }\textbf {\bibinfo {volume} {128}},\ \bibinfo {pages}
  {090602} (\bibinfo {year} {2022})}\BibitemShut {NoStop}%
\bibitem [{\citenamefont {Micadei}\ \emph {et~al.}(2019)\citenamefont
  {Micadei}, \citenamefont {Peterson}, \citenamefont {Souza}, \citenamefont
  {Sarthour}, \citenamefont {Oliveira}, \citenamefont {Landi}, \citenamefont
  {Batalh{\~a}o}, \citenamefont {Serra},\ and\ \citenamefont
  {Lutz}}]{Micadei2019}%
  \BibitemOpen
  \bibfield  {author} {\bibinfo {author} {\bibfnamefont {K.}~\bibnamefont
  {Micadei}}, \bibinfo {author} {\bibfnamefont {J.~P.~S.}\ \bibnamefont
  {Peterson}}, \bibinfo {author} {\bibfnamefont {A.~M.}\ \bibnamefont {Souza}},
  \bibinfo {author} {\bibfnamefont {R.~S.}\ \bibnamefont {Sarthour}}, \bibinfo
  {author} {\bibfnamefont {I.~S.}\ \bibnamefont {Oliveira}}, \bibinfo {author}
  {\bibfnamefont {G.~T.}\ \bibnamefont {Landi}}, \bibinfo {author}
  {\bibfnamefont {T.~B.}\ \bibnamefont {Batalh{\~a}o}}, \bibinfo {author}
  {\bibfnamefont {R.~M.}\ \bibnamefont {Serra}}, \ and\ \bibinfo {author}
  {\bibfnamefont {E.}~\bibnamefont {Lutz}},\ }\href {\doibase
  10.1038/s41467-019-10333-7} {\bibfield  {journal} {\bibinfo  {journal} {Nat.
  Commun.}\ }\textbf {\bibinfo {volume} {10}},\ \bibinfo {pages} {2456}
  (\bibinfo {year} {2019})}\BibitemShut {NoStop}%
\bibitem [{\citenamefont {Brunner}\ \emph {et~al.}(2014)\citenamefont
  {Brunner}, \citenamefont {Huber}, \citenamefont {Linden}, \citenamefont
  {Popescu}, \citenamefont {Silva},\ and\ \citenamefont
  {Skrzypczyk}}]{PhysRevE.89.032115}%
  \BibitemOpen
  \bibfield  {author} {\bibinfo {author} {\bibfnamefont {N.}~\bibnamefont
  {Brunner}}, \bibinfo {author} {\bibfnamefont {M.}~\bibnamefont {Huber}},
  \bibinfo {author} {\bibfnamefont {N.}~\bibnamefont {Linden}}, \bibinfo
  {author} {\bibfnamefont {S.}~\bibnamefont {Popescu}}, \bibinfo {author}
  {\bibfnamefont {R.}~\bibnamefont {Silva}}, \ and\ \bibinfo {author}
  {\bibfnamefont {P.}~\bibnamefont {Skrzypczyk}},\ }\href {\doibase
  10.1103/PhysRevE.89.032115} {\bibfield  {journal} {\bibinfo  {journal} {Phys.
  Rev. E}\ }\textbf {\bibinfo {volume} {89}},\ \bibinfo {pages} {032115}
  (\bibinfo {year} {2014})}\BibitemShut {NoStop}%
\bibitem [{\citenamefont {Curzon}\ and\ \citenamefont
  {Ahlborn}(1975)}]{10.1119/1.10023}%
  \BibitemOpen
  \bibfield  {author} {\bibinfo {author} {\bibfnamefont {F.~L.}\ \bibnamefont
  {Curzon}}\ and\ \bibinfo {author} {\bibfnamefont {B.}~\bibnamefont
  {Ahlborn}},\ }\href {\doibase 10.1119/1.10023} {\bibfield  {journal}
  {\bibinfo  {journal} {Am. J. Phys.}\ }\textbf {\bibinfo {volume} {43}},\
  \bibinfo {pages} {22} (\bibinfo {year} {1975})}\BibitemShut {NoStop}%
\bibitem [{\citenamefont {Allahverdyan}\ \emph {et~al.}(2010)\citenamefont
  {Allahverdyan}, \citenamefont {Hovhannisyan},\ and\ \citenamefont
  {Mahler}}]{PhysRevE.81.051129}%
  \BibitemOpen
  \bibfield  {author} {\bibinfo {author} {\bibfnamefont {A.~E.}\ \bibnamefont
  {Allahverdyan}}, \bibinfo {author} {\bibfnamefont {K.}~\bibnamefont
  {Hovhannisyan}}, \ and\ \bibinfo {author} {\bibfnamefont {G.}~\bibnamefont
  {Mahler}},\ }\href {\doibase 10.1103/PhysRevE.81.051129} {\bibfield
  {journal} {\bibinfo  {journal} {Phys. Rev. E}\ }\textbf {\bibinfo {volume}
  {81}},\ \bibinfo {pages} {051129} (\bibinfo {year} {2010})}\BibitemShut
  {NoStop}%
\bibitem [{\citenamefont {Scully}\ \emph {et~al.}(2003)\citenamefont {Scully},
  \citenamefont {Zubairy}, \citenamefont {Agarwal},\ and\ \citenamefont
  {Walther}}]{doi:10.1126/science.1078955}%
  \BibitemOpen
  \bibfield  {author} {\bibinfo {author} {\bibfnamefont {M.~O.}\ \bibnamefont
  {Scully}}, \bibinfo {author} {\bibfnamefont {M.~S.}\ \bibnamefont {Zubairy}},
  \bibinfo {author} {\bibfnamefont {G.~S.}\ \bibnamefont {Agarwal}}, \ and\
  \bibinfo {author} {\bibfnamefont {H.}~\bibnamefont {Walther}},\ }\href
  {\doibase 10.1126/science.1078955} {\bibfield  {journal} {\bibinfo  {journal}
  {Science}\ }\textbf {\bibinfo {volume} {299}},\ \bibinfo {pages} {862}
  (\bibinfo {year} {2003})}\BibitemShut {NoStop}%
\bibitem [{\citenamefont {Geva}\ and\ \citenamefont
  {Kosloff}(1992)}]{geva1992quantum}%
  \BibitemOpen
  \bibfield  {author} {\bibinfo {author} {\bibfnamefont {E.}~\bibnamefont
  {Geva}}\ and\ \bibinfo {author} {\bibfnamefont {R.}~\bibnamefont {Kosloff}},\
  }\href@noop {} {\bibfield  {journal} {\bibinfo  {journal} {J. Chem. Phys.}\
  }\textbf {\bibinfo {volume} {96}},\ \bibinfo {pages} {3054} (\bibinfo {year}
  {1992})}\BibitemShut {NoStop}%
\bibitem [{\citenamefont {Correa}\ \emph {et~al.}(2013)\citenamefont {Correa},
  \citenamefont {Palao}, \citenamefont {Alonso},\ and\ \citenamefont
  {Adesso}}]{Correa2013QuantumenhancedAR}%
  \BibitemOpen
  \bibfield  {author} {\bibinfo {author} {\bibfnamefont {L.~A.}\ \bibnamefont
  {Correa}}, \bibinfo {author} {\bibfnamefont {J.~P.}\ \bibnamefont {Palao}},
  \bibinfo {author} {\bibfnamefont {D.}~\bibnamefont {Alonso}}, \ and\ \bibinfo
  {author} {\bibfnamefont {G.}~\bibnamefont {Adesso}},\ }\href
  {https://api.semanticscholar.org/CorpusID:18645006} {\bibfield  {journal}
  {\bibinfo  {journal} {Sci. Rep.}\ }\textbf {\bibinfo {volume} {4}} (\bibinfo
  {year} {2013})}\BibitemShut {NoStop}%
\bibitem [{\citenamefont {Schulman}\ and\ \citenamefont
  {Vazirani}(1999)}]{schulman1999}%
  \BibitemOpen
  \bibfield  {author} {\bibinfo {author} {\bibfnamefont {L.~J.}\ \bibnamefont
  {Schulman}}\ and\ \bibinfo {author} {\bibfnamefont {U.~V.}\ \bibnamefont
  {Vazirani}},\ }in\ \href@noop {} {\emph {\bibinfo {booktitle} {Proceedings of
  the thirty-first annual ACM symposium on Theory of computing}}}\ (\bibinfo
  {year} {1999})\ pp.\ \bibinfo {pages} {322--329}\BibitemShut {NoStop}%
\bibitem [{\citenamefont {Fernandez}\ \emph {et~al.}(2004)\citenamefont
  {Fernandez}, \citenamefont {Lloyd}, \citenamefont {Mor},\ and\ \citenamefont
  {Roychowdhury}}]{fernandez2004}%
  \BibitemOpen
  \bibfield  {author} {\bibinfo {author} {\bibfnamefont {J.~M.}\ \bibnamefont
  {Fernandez}}, \bibinfo {author} {\bibfnamefont {S.}~\bibnamefont {Lloyd}},
  \bibinfo {author} {\bibfnamefont {T.}~\bibnamefont {Mor}}, \ and\ \bibinfo
  {author} {\bibfnamefont {V.}~\bibnamefont {Roychowdhury}},\ }\href {\doibase
  10.1142/S0219749904000419} {\bibfield  {journal} {\bibinfo  {journal} {Int.
  J. Quantum Inf.}\ }\textbf {\bibinfo {volume} {02}},\ \bibinfo {pages} {461}
  (\bibinfo {year} {2004})}\BibitemShut {NoStop}%
\bibitem [{\citenamefont {Boykin}\ \emph {et~al.}(2002)\citenamefont {Boykin},
  \citenamefont {Mor}, \citenamefont {Roychowdhury}, \citenamefont {Vatan},\
  and\ \citenamefont {Vrijen}}]{Boykin2002}%
  \BibitemOpen
  \bibfield  {author} {\bibinfo {author} {\bibfnamefont {P.~O.}\ \bibnamefont
  {Boykin}}, \bibinfo {author} {\bibfnamefont {T.}~\bibnamefont {Mor}},
  \bibinfo {author} {\bibfnamefont {V.}~\bibnamefont {Roychowdhury}}, \bibinfo
  {author} {\bibfnamefont {F.}~\bibnamefont {Vatan}}, \ and\ \bibinfo {author}
  {\bibfnamefont {R.}~\bibnamefont {Vrijen}},\ }\href {\doibase
  10.1073/pnas.241641898} {\bibfield  {journal} {\bibinfo  {journal} {Proc.
  Natl. Acad. Sci. U.S.A.}\ }\textbf {\bibinfo {volume} {99}},\ \bibinfo
  {pages} {3388} (\bibinfo {year} {2002})}\BibitemShut {NoStop}%
\bibitem [{\citenamefont {Rodríguez-Briones}\ \emph
  {et~al.}(2017)\citenamefont {Rodríguez-Briones}, \citenamefont {Li},
  \citenamefont {Peng}, \citenamefont {Mor}, \citenamefont {Weinstein},\ and\
  \citenamefont {Laflamme}}]{Rod2017}%
  \BibitemOpen
  \bibfield  {author} {\bibinfo {author} {\bibfnamefont {N.~A.}\ \bibnamefont
  {Rodríguez-Briones}}, \bibinfo {author} {\bibfnamefont {J.}~\bibnamefont
  {Li}}, \bibinfo {author} {\bibfnamefont {X.}~\bibnamefont {Peng}}, \bibinfo
  {author} {\bibfnamefont {T.}~\bibnamefont {Mor}}, \bibinfo {author}
  {\bibfnamefont {Y.}~\bibnamefont {Weinstein}}, \ and\ \bibinfo {author}
  {\bibfnamefont {R.}~\bibnamefont {Laflamme}},\ }\href {\doibase
  10.1088/1367-2630/aa8fe0} {\bibfield  {journal} {\bibinfo  {journal} {New J.
  Phys.}\ }\textbf {\bibinfo {volume} {19}},\ \bibinfo {pages} {113047}
  (\bibinfo {year} {2017})}\BibitemShut {NoStop}%
\bibitem [{\citenamefont {Rodr\'{\i}guez-Briones}\ \emph
  {et~al.}(2017)\citenamefont {Rodr\'{\i}guez-Briones}, \citenamefont
  {Mart\'{\i}n-Mart\'{\i}nez}, \citenamefont {Kempf},\ and\ \citenamefont
  {Laflamme}}]{PhysRevLett.119.050502}%
  \BibitemOpen
  \bibfield  {author} {\bibinfo {author} {\bibfnamefont {N.~A.}\ \bibnamefont
  {Rodr\'{\i}guez-Briones}}, \bibinfo {author} {\bibfnamefont {E.}~\bibnamefont
  {Mart\'{\i}n-Mart\'{\i}nez}}, \bibinfo {author} {\bibfnamefont
  {A.}~\bibnamefont {Kempf}}, \ and\ \bibinfo {author} {\bibfnamefont
  {R.}~\bibnamefont {Laflamme}},\ }\href {\doibase
  10.1103/PhysRevLett.119.050502} {\bibfield  {journal} {\bibinfo  {journal}
  {Phys. Rev. Lett.}\ }\textbf {\bibinfo {volume} {119}},\ \bibinfo {pages}
  {050502} (\bibinfo {year} {2017})}\BibitemShut {NoStop}%
\bibitem [{\citenamefont {Linden}\ \emph {et~al.}(2010)\citenamefont {Linden},
  \citenamefont {Popescu},\ and\ \citenamefont
  {Skrzypczyk}}]{PhysRevLett.105.130401}%
  \BibitemOpen
  \bibfield  {author} {\bibinfo {author} {\bibfnamefont {N.}~\bibnamefont
  {Linden}}, \bibinfo {author} {\bibfnamefont {S.}~\bibnamefont {Popescu}}, \
  and\ \bibinfo {author} {\bibfnamefont {P.}~\bibnamefont {Skrzypczyk}},\
  }\href {\doibase 10.1103/PhysRevLett.105.130401} {\bibfield  {journal}
  {\bibinfo  {journal} {Phys. Rev. Lett.}\ }\textbf {\bibinfo {volume} {105}},\
  \bibinfo {pages} {130401} (\bibinfo {year} {2010})}\BibitemShut {NoStop}%
\bibitem [{\citenamefont {Venturelli}\ \emph {et~al.}(2013)\citenamefont
  {Venturelli}, \citenamefont {Fazio},\ and\ \citenamefont
  {Giovannetti}}]{PhysRevLett.110.256801}%
  \BibitemOpen
  \bibfield  {author} {\bibinfo {author} {\bibfnamefont {D.}~\bibnamefont
  {Venturelli}}, \bibinfo {author} {\bibfnamefont {R.}~\bibnamefont {Fazio}}, \
  and\ \bibinfo {author} {\bibfnamefont {V.}~\bibnamefont {Giovannetti}},\
  }\href {\doibase 10.1103/PhysRevLett.110.256801} {\bibfield  {journal}
  {\bibinfo  {journal} {Phys. Rev. Lett.}\ }\textbf {\bibinfo {volume} {110}},\
  \bibinfo {pages} {256801} (\bibinfo {year} {2013})}\BibitemShut {NoStop}%
\bibitem [{\citenamefont {Baugh}\ \emph {et~al.}(2005)\citenamefont {Baugh},
  \citenamefont {Moussa}, \citenamefont {Ryan}, \citenamefont {Nayak},\ and\
  \citenamefont {Laflamme}}]{Baugh2005}%
  \BibitemOpen
  \bibfield  {author} {\bibinfo {author} {\bibfnamefont {J.}~\bibnamefont
  {Baugh}}, \bibinfo {author} {\bibfnamefont {O.}~\bibnamefont {Moussa}},
  \bibinfo {author} {\bibfnamefont {C.~A.}\ \bibnamefont {Ryan}}, \bibinfo
  {author} {\bibfnamefont {A.}~\bibnamefont {Nayak}}, \ and\ \bibinfo {author}
  {\bibfnamefont {R.}~\bibnamefont {Laflamme}},\ }\href {\doibase
  10.1038/nature04272} {\bibfield  {journal} {\bibinfo  {journal} {Nature}\
  }\textbf {\bibinfo {volume} {438}},\ \bibinfo {pages} {470} (\bibinfo {year}
  {2005})}\BibitemShut {NoStop}%
\bibitem [{\citenamefont {Ryan}\ \emph {et~al.}(2008)\citenamefont {Ryan},
  \citenamefont {Moussa}, \citenamefont {Baugh},\ and\ \citenamefont
  {Laflamme}}]{PhysRevLett.100.140501}%
  \BibitemOpen
  \bibfield  {author} {\bibinfo {author} {\bibfnamefont {C.~A.}\ \bibnamefont
  {Ryan}}, \bibinfo {author} {\bibfnamefont {O.}~\bibnamefont {Moussa}},
  \bibinfo {author} {\bibfnamefont {J.}~\bibnamefont {Baugh}}, \ and\ \bibinfo
  {author} {\bibfnamefont {R.}~\bibnamefont {Laflamme}},\ }\href {\doibase
  10.1103/PhysRevLett.100.140501} {\bibfield  {journal} {\bibinfo  {journal}
  {Phys. Rev. Lett.}\ }\textbf {\bibinfo {volume} {100}},\ \bibinfo {pages}
  {140501} (\bibinfo {year} {2008})}\BibitemShut {NoStop}%
\bibitem [{\citenamefont {Soldati}\ \emph {et~al.}(2022)\citenamefont
  {Soldati}, \citenamefont {Dasari}, \citenamefont {Wrachtrup},\ and\
  \citenamefont {Lutz}}]{PhysRevLett.129.030601}%
  \BibitemOpen
  \bibfield  {author} {\bibinfo {author} {\bibfnamefont {R.~R.}\ \bibnamefont
  {Soldati}}, \bibinfo {author} {\bibfnamefont {D.~B.~R.}\ \bibnamefont
  {Dasari}}, \bibinfo {author} {\bibfnamefont {J.}~\bibnamefont {Wrachtrup}}, \
  and\ \bibinfo {author} {\bibfnamefont {E.}~\bibnamefont {Lutz}},\ }\href
  {\doibase 10.1103/PhysRevLett.129.030601} {\bibfield  {journal} {\bibinfo
  {journal} {Phys. Rev. Lett.}\ }\textbf {\bibinfo {volume} {129}},\ \bibinfo
  {pages} {030601} (\bibinfo {year} {2022})}\BibitemShut {NoStop}%
\bibitem [{\citenamefont {Long}\ \emph {et~al.}(2022)\citenamefont {Long},
  \citenamefont {He}, \citenamefont {Zhang}, \citenamefont {Tang},
  \citenamefont {Lin}, \citenamefont {Liu}, \citenamefont {Nie}, \citenamefont
  {Feng}, \citenamefont {Li}, \citenamefont {Xin}, \citenamefont {Ai},\ and\
  \citenamefont {Lu}}]{PhysRevLett.129.070502}%
  \BibitemOpen
  \bibfield  {author} {\bibinfo {author} {\bibfnamefont {X.}~\bibnamefont
  {Long}}, \bibinfo {author} {\bibfnamefont {W.-T.}\ \bibnamefont {He}},
  \bibinfo {author} {\bibfnamefont {N.-N.}\ \bibnamefont {Zhang}}, \bibinfo
  {author} {\bibfnamefont {K.}~\bibnamefont {Tang}}, \bibinfo {author}
  {\bibfnamefont {Z.}~\bibnamefont {Lin}}, \bibinfo {author} {\bibfnamefont
  {H.}~\bibnamefont {Liu}}, \bibinfo {author} {\bibfnamefont {X.}~\bibnamefont
  {Nie}}, \bibinfo {author} {\bibfnamefont {G.}~\bibnamefont {Feng}}, \bibinfo
  {author} {\bibfnamefont {J.}~\bibnamefont {Li}}, \bibinfo {author}
  {\bibfnamefont {T.}~\bibnamefont {Xin}}, \bibinfo {author} {\bibfnamefont
  {Q.}~\bibnamefont {Ai}}, \ and\ \bibinfo {author} {\bibfnamefont
  {D.}~\bibnamefont {Lu}},\ }\href {\doibase 10.1103/PhysRevLett.129.070502}
  {\bibfield  {journal} {\bibinfo  {journal} {Phys. Rev. Lett.}\ }\textbf
  {\bibinfo {volume} {129}},\ \bibinfo {pages} {070502} (\bibinfo {year}
  {2022})}\BibitemShut {NoStop}%
\bibitem [{\citenamefont {Xin}\ \emph {et~al.}(2021)\citenamefont {Xin},
  \citenamefont {Che}, \citenamefont {Xi}, \citenamefont {Singh}, \citenamefont
  {Nie}, \citenamefont {Li}, \citenamefont {Dong},\ and\ \citenamefont
  {Lu}}]{PhysRevLett.126.110502}%
  \BibitemOpen
  \bibfield  {author} {\bibinfo {author} {\bibfnamefont {T.}~\bibnamefont
  {Xin}}, \bibinfo {author} {\bibfnamefont {L.}~\bibnamefont {Che}}, \bibinfo
  {author} {\bibfnamefont {C.}~\bibnamefont {Xi}}, \bibinfo {author}
  {\bibfnamefont {A.}~\bibnamefont {Singh}}, \bibinfo {author} {\bibfnamefont
  {X.}~\bibnamefont {Nie}}, \bibinfo {author} {\bibfnamefont {J.}~\bibnamefont
  {Li}}, \bibinfo {author} {\bibfnamefont {Y.}~\bibnamefont {Dong}}, \ and\
  \bibinfo {author} {\bibfnamefont {D.}~\bibnamefont {Lu}},\ }\href {\doibase
  10.1103/PhysRevLett.126.110502} {\bibfield  {journal} {\bibinfo  {journal}
  {Phys. Rev. Lett.}\ }\textbf {\bibinfo {volume} {126}},\ \bibinfo {pages}
  {110502} (\bibinfo {year} {2021})}\BibitemShut {NoStop}%
\bibitem [{sup()}]{supply}%
  \BibitemOpen
  \href@noop {} {}\bibinfo {note} {See Supplemental Information for more
  details.}\BibitemShut {Stop}%
\bibitem [{\citenamefont {Tseng}\ \emph {et~al.}(1999)\citenamefont {Tseng},
  \citenamefont {Somaroo}, \citenamefont {Sharf}, \citenamefont {Knill},
  \citenamefont {Laflamme}, \citenamefont {Havel},\ and\ \citenamefont
  {Cory}}]{PhysRevA.61.012302}%
  \BibitemOpen
  \bibfield  {author} {\bibinfo {author} {\bibfnamefont {C.~H.}\ \bibnamefont
  {Tseng}}, \bibinfo {author} {\bibfnamefont {S.}~\bibnamefont {Somaroo}},
  \bibinfo {author} {\bibfnamefont {Y.}~\bibnamefont {Sharf}}, \bibinfo
  {author} {\bibfnamefont {E.}~\bibnamefont {Knill}}, \bibinfo {author}
  {\bibfnamefont {R.}~\bibnamefont {Laflamme}}, \bibinfo {author}
  {\bibfnamefont {T.~F.}\ \bibnamefont {Havel}}, \ and\ \bibinfo {author}
  {\bibfnamefont {D.~G.}\ \bibnamefont {Cory}},\ }\href {\doibase
  10.1103/PhysRevA.61.012302} {\bibfield  {journal} {\bibinfo  {journal} {Phys.
  Rev. A}\ }\textbf {\bibinfo {volume} {61}},\ \bibinfo {pages} {012302}
  (\bibinfo {year} {1999})}\BibitemShut {NoStop}%
\bibitem [{\citenamefont {Uzdin}\ \emph {et~al.}(2015)\citenamefont {Uzdin},
  \citenamefont {Levy},\ and\ \citenamefont {Kosloff}}]{PhysRevX.5.031044}%
  \BibitemOpen
  \bibfield  {author} {\bibinfo {author} {\bibfnamefont {R.}~\bibnamefont
  {Uzdin}}, \bibinfo {author} {\bibfnamefont {A.}~\bibnamefont {Levy}}, \ and\
  \bibinfo {author} {\bibfnamefont {R.}~\bibnamefont {Kosloff}},\ }\href
  {\doibase 10.1103/PhysRevX.5.031044} {\bibfield  {journal} {\bibinfo
  {journal} {Phys. Rev. X}\ }\textbf {\bibinfo {volume} {5}},\ \bibinfo {pages}
  {031044} (\bibinfo {year} {2015})}\BibitemShut {NoStop}%
\bibitem [{\citenamefont {Callen}(1991)}]{callen1991thermodynamics}%
  \BibitemOpen
  \bibfield  {author} {\bibinfo {author} {\bibfnamefont {H.~B.}\ \bibnamefont
  {Callen}},\ }\href@noop {} {\emph {\bibinfo {title} {Thermodynamics and an
  Introduction to Thermostatistics}}}\ (\bibinfo  {publisher} {John Wiley \&
  Sons},\ \bibinfo {year} {1991})\BibitemShut {NoStop}%
\bibitem [{\citenamefont {Levy}\ and\ \citenamefont
  {Kosloff}(2012)}]{PhysRevLett.108.070604}%
  \BibitemOpen
  \bibfield  {author} {\bibinfo {author} {\bibfnamefont {A.}~\bibnamefont
  {Levy}}\ and\ \bibinfo {author} {\bibfnamefont {R.}~\bibnamefont {Kosloff}},\
  }\href {\doibase 10.1103/PhysRevLett.108.070604} {\bibfield  {journal}
  {\bibinfo  {journal} {Phys. Rev. Lett.}\ }\textbf {\bibinfo {volume} {108}},\
  \bibinfo {pages} {070604} (\bibinfo {year} {2012})}\BibitemShut {NoStop}%
\bibitem [{\citenamefont {Rodr\'{\i}guez-Briones}\ and\ \citenamefont
  {Laflamme}(2016)}]{PhysRevLett.116.170501}%
  \BibitemOpen
  \bibfield  {author} {\bibinfo {author} {\bibfnamefont {N.~A.}\ \bibnamefont
  {Rodr\'{\i}guez-Briones}}\ and\ \bibinfo {author} {\bibfnamefont
  {R.}~\bibnamefont {Laflamme}},\ }\href {\doibase
  10.1103/PhysRevLett.116.170501} {\bibfield  {journal} {\bibinfo  {journal}
  {Phys. Rev. Lett.}\ }\textbf {\bibinfo {volume} {116}},\ \bibinfo {pages}
  {170501} (\bibinfo {year} {2016})}\BibitemShut {NoStop}%
\end{thebibliography}
\end{document}